\documentclass[twocolumn, numberedappendix, nofootinbib]{revtex4-2}

\usepackage{graphicx}
\usepackage{dcolumn}
\usepackage{amssymb, amsmath}
\usepackage{siunitx}

\usepackage[colorlinks=true, 
            linkcolor=blue, 
            urlcolor=blue, 
            citecolor=magenta]{hyperref}

\usepackage{orcidlink}

\usepackage{xspace}
\def\Sref#1{Sec.~\ref{#1}\xspace}
\def\Fref#1{Fig.~\ref{#1}\xspace}
\def\Eref#1{Eq.~\eqref{#1}\xspace}

\begin{document}

\title{Beyond collective fluctuations:\\probing micro-image swarms in lensed quasars with intensity interferometry}

\author{Ashish Kumar Meena\orcidlink{0000-0002-7876-4321}}
\email{akm@iisc.ac.in}
\affiliation{Department of Physics, Indian Institute of Science, Bengaluru 560012, India}

\author{Prasenjit Saha\orcidlink{0000-0003-0136-2153}}
\email{psaha@physik.uzh.ch}
\affiliation{Physik-Institut, University of Zurich, Winterthurerstrasse 190, CH-8057 Zurich, Switzerland}

\begin{abstract}
Each strongly lensed image of a quasar behind a lensing galaxy (or galaxy cluster) is composed of a swarm of micro-images. This is a result of microlensing due to stellar-scale substructure in the lens. The presence of microlenses forms a network of micro-caustics, and a source transiting these micro-caustics gives rise to variation in observed strongly lensed images. These micro-image swarms are currently observable only through collective intensity fluctuations, which hide the underlying information on the stellar (and compact dark matter, if any) mass functions within the swarm. To unlock the information present in micro-image swarms, it is necessary to explore new techniques. In this work, we study the prospects of determining the micro-image swarm size in lensed quasar images using the intensity interferometry~(i.e., the Hanbury Brown \& Twiss effect). We consider QSO~2237+0305 and PS~J0147+4630, two of the brightest lens quasars in the sky, and study micro-image swarm features in visibility space for both macro-minimum and macro-saddle-point images. At the end, we argue that, with ongoing and expected technical advances, observations of micro-image swarms are plausible, at least for the brightest lensed quasars. 
\end{abstract}

\maketitle

\section{Introduction}
\label{sec:intro}
Gravitational lensing refers to the bending of light as it passes close to an intermediate matter distribution between observer and light source~\citep[e.g.,][]{1992grle.book.....S}. Lensing within our own Galaxy and of distant sources has allowed us to study and constrain various properties of the Universe. For example, microlensing of stars within our Galaxy and M31 has provided stringent constraints on the nature of dark matter~\citep[e.g.,][]{2007A&A...469..387T, 2019NatAs...3..524N, 2025ApJS..280...49M}. Strong lensing of distant quasars and supernovae has allowed us to measure the Hubble constant with a few percent precision~\cite[e.g.,][]{2020MNRAS.498.1420W, 2025ApJ...979...13P, 2025arXiv250912319S}. In addition, microlensing within strongly lensed images allows us to probe stellar and dark population within lensing galaxies and galaxy clusters~\citep[e.g.,][]{2017ApJ...836L..18M, 2018PhRvD..97b3518O}. Recently, combined strong and microlensing has even enabled us to observe individual stars at the cosmic noon and beyond~\citep[e.g.,][]{2018NatAs...2..334K, 2022Natur.603..815W, 2023ApJ...944L...6M}. In strong lensing by galaxies or galaxy clusters, the typical image separation is of the order of an arcsecond to tens of arcseconds, which can be comfortably resolved with one-meter-class optical telescopes. Currently, state-of-the-art optical and near-infrared) telescopes, such as the Hubble Space Telescope (HST) and the James Webb Space Telescope (JWST), reach resolutions of tens of milli-arcseconds. Even higher resolutions, i.e., (sub-)milli-arcseconds, have been achieved at radio wavelengths using very long~(Earth-size) baselines and have been used to search for dark matter subhalos at~$\sim10^7-10^8~{\rm M_\odot}$~\citep[e.g.,][]{2018MNRAS.478.4816S, 2019MNRAS.485.3009H, 2020MNRAS.492..742A}.

Achieving (sub-)milli-arcseconds resolution at optical wavelengths (too small for standard optical telescopes) allows us to resolve nearby bright individual stars. The first such observation was made a century ago by~\citet{1921ApJ....53..249M} who used the Michelson stellar interferometer with a baseline of six meters to measure the (angular) diameter of $\alpha$-Orionis~(i.e., Betelgeuse). In recent times, with Michelson stellar interferometry, the CHARA array with the maximum baseline of~$\sim300$~meters, has measured sizes of more than a hundred nearby stars~\citep[e.g.,][]{2012ApJ...746..101B, 2012ApJ...757..112B, 2015MNRAS.447..846B}, including observations of star-spots~\citep{2016Natur.533..217R}. 

In addition to Michelson interferometry, another technique for measuring the sizes of individual stars is \textit{Intensity Interferometry}~(II)~\cite{1974iiaa.book.....B}, which measures the temporal correlation between photons at two receivers. This technique was pioneered by Hanbury Brown \& Twiss and is also known as the HBT~effect. In II, when the two telescopes are close together, their signals are correlated, and as the separation increases, the correlation decreases, allowing us to measure the spatial source properties. \citet{1952Natur.170.1061H} demonstrated the first use of II by measuring the size of Cygnus and Cassiopeia at radio wavelengths. Following this feat, in 1974, the Narrabri Stellar Intensity Interferometer (NSII) measured the diameters of 32 stars at optical wavelengths~\cite{1974MNRAS.167..121H}. In recent years, II has again been used to measure the diameters of various stars using Cherenkov~\cite{2024MNRAS.529.4387A, 2025ApJ...995..191A, 2025MNRAS.537.2334V} and optical~\citep{2018MNRAS.480..245G, 2020MNRAS.494..218R, 2023AJ....165..117M} telescopes.

At present, the most straightforward and promising use of II is to study various properties of bright stars and transients~\cite{2012NewAR..56..143D}, as the current and upcoming Cherenkov telescopes can only target objects brighter than~$\sim8-10$~AB magnitudes before hitting the night sky background~\citep{Saha:2025uG}. However, keeping in mind the ongoing technological advances, multiple new science cases for II, including cosmological applications, have also been proposed recently. These include prospects for resolving gravitational wave binaries~\cite{2020MNRAS.498.4577B}, resolving central regions of active galactic nuclei~(AGN) using baselines of tens of kilometers and constraining the Hubble constant~($H_0$)~\cite{2024PhRvD.109l3029D}, separating the intrinsic and microlensing-induced variability of the brightest lensed quasar(s), ultimately improving~$H_0$ measurements~\cite{2025arXiv251212470S}, and addressing the nature of dark matter by studying microlensing-induced astrometric jitter in the brightest lensed quasar(s)~\cite{2026arXiv260212717K}.

In quasars, at optical wavelengths, the dominant contribution to the observed flux comes from the accretion disk whose size is $\sim10^{-3}$~parsec, and at cosmological distances, the corresponding angular size is~$\lesssim10^{-7}$~arcseconds. Such objects, even at optical wavelengths, would need baselines of thousands of kilometers to be resolved. However, in lensed quasars, the presence of microlenses splits the strongly lensed image further into micro-images with separations~$\gtrsim10^{-6}$~arcseconds. A preliminary calculation\footnote{C.~Magnoli (2022) Bachelor project, University of Zurich.} noted that the size of the swarm of these micro-images can be measured with baselines spanning tens of kilometers, which is similar to the separation between the Very Large Telescope~(VLT) and the Extremely Large Telescope~\citep[ELT;][]{2023ConPh..64...47P} in Chile. However, one question still remains: \textit{Is the photon flux from the brightest lensed quasars sufficient to be observed and studied with II?} To address this question, in our work, we consider two of the brightest lensed quasars, i.e., QSO~2237+0305 and QSO~J0147+4630, to study the observed photon flux for these sources and the effect of microlensing in the visibility space. For both of these lensed quasars, we consider one minimum and one saddle-point strongly lensed image to investigate microlensing effects on the observed photon flux. For both of these lensed quasars, the brightest lensed image has an apparent magnitude of~$\sim16-17$~AB, which is too faint for the current II facilities. Hence, we also examine the feasibility of such observations in the future.

The current work is organized as follows. In \Sref{sec:basic_gl} and \Sref{sec:basic_II}, we discuss the relevant basics of gravitational lensing and II. In \Sref{sec:pml} and \Sref{sec:crl}, we study microlensing effects on observed photon correlation in the (complex) visibility space for an isolated point lens and a point lens with external effects, respectively. \Sref{sec:cross} and \Sref{sec:psj0147} study the formation of micro-image swarm and the effect of microlensing on observed photon correlation in strongly lensed images for QSO~2237+0305 and QSO~J0147+4630 with II, respectively. In \Sref{sec:strategy}, the feasibility of such observations and possible future observing strategies to resolve micro-image swarms is discussed. We conclude our work in \ref{sec:conclusion}. Throughout this work, we use flat $\Lambda$CDM cosmology with parameters,~$H_0=70\,{\rm km\,s^{-1}\,Mpc^{-1}}$, $\Omega_{m,0}=0.3$ and~$\Omega_{\Lambda}=0.7$. Unless mentioned otherwise, following QSO~2237+0305, we fix the lens and source redshifts to~$z_d=0.0394$ and~$z_s=1.695$, respectively. For the above lens redshift, the angular diameter distance is 160.89~Mpc and $1''=0.78~{\rm Kpc}$. The brightest image for QSO~2237+0305 has an apparent magnitude of~$16.73$~AB in G-band\footnote{\url{https://research.ast.cam.ac.uk/lensedquasars/indiv/Q2237+030.html}}. The corresponding lensing-corrected magnitude is 18.43~AB (see \ref{sec:cross} for more details) and is used as our default source magnitude. Throughout this work, for simplicity, we also assume a point source. The implications of this assumption are briefly discussed in \Sref{sec:strategy}.

\section{Lensing Basics}
\label{sec:basic_gl}
In gravitational lensing, one of the fundamental quantities is the lens equation, which relates the observed angular image position~($\pmb{\theta}$) and the corresponding unlensed angular source position~($\pmb{\beta}$), and is given as~\cite{1992grle.book.....S, 1996astro.ph..6001N},
\begin{equation}
    \pmb{\beta} = \pmb{\theta} - \pmb{\nabla}\psi(\pmb{\theta}),
    \label{eq:lens_eq}
\end{equation}
where~$\psi(\pmb{\theta})$ is the 2D projected lens potential. Often, in gravitational lensing, the background source is multiply imaged, a regime known as \textit{strong lensing}. The magnification factor for a lensed image, assuming a point source, is given as,
\begin{equation}
    \mu(\pmb{\theta}) = \left|\frac{\partial \pmb{\beta}}{\partial\pmb{\theta}}\right|^{-1} = \frac{1}{\mu_t\:\mu_r}=\frac{1}{(1-\kappa-\gamma)(1-\kappa+\gamma)},
    \label{eq:mu}
\end{equation}
where~$\kappa$ and~$\gamma$ represent the well-known convergence and shear values at the image position, respectively. $\mu_t$ and $\mu_r$ represent the tangential and radial magnification components\footnote{Typically, a large tangential~(radial) magnification implies that lensed image would be stretched in tangential~(radial) direction with respect to the lens center.}. As we can see from the equation, the convergence~(shear) governs the isotropic~(directional) distortion in the observed image(s). For our current work, the above two equations are sufficient as we are only interested in image positions and their magnification factors as discussed below.

\newcommand{\pcoh}{\Phi}
\newcommand{\spcoh}{\Delta\tau}
\newcommand{\tobs}{T}
\newcommand{\Aeff}{A_{\rm eff}}
\newcommand{\Nchan}{N_{\rm ch}}

\section{Essentials of Intensity Interferometry}
\label{sec:basic_II}

\begin{figure}
\centering
\includegraphics[width=\linewidth]{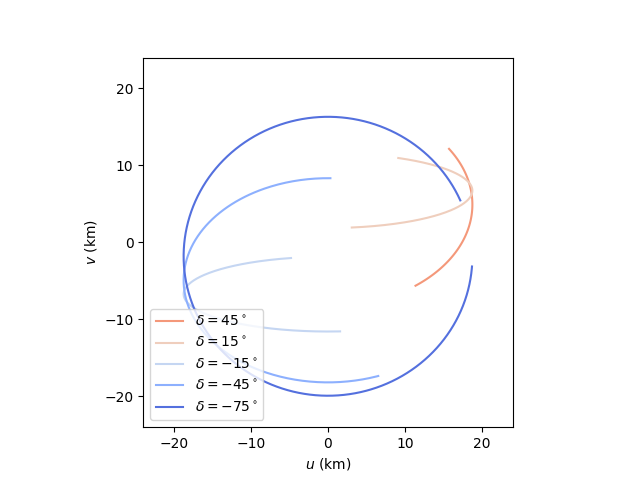}
\caption{An illustration of tracks in the interferometric $(u,v)$ plane.  Assumed here is a baseline of two telescopes at latitude about $25^\circ\,$S, separated by $20\,$km in the ENE--WSW direction. The tracks represent the baseline projected on a plane perpendicular to the line of sight, as the source moves across the sky. The start and endpoints of each track correspond to the source rising and setting.}
\label{fig:uvtracks}
\end{figure}

In astronomical interferometry, the fundamental observable is the so-called complex visibility, which is the spatial Fourier transform of the source brightness distribution.  For micro-images within a single lensed macro-image, the complex visibility is,
\begin{equation}
  V(u, v) \propto \sum_j \big| \mu^{(j)} \big|
  \exp \left[ \frac{2\pi i}{\lambda}
              \left(u\theta^{(j)}_x + v\theta^{(j)}_y\right) \right],
\end{equation}
normalized as $V(0,0)=1$.  Here, $\theta^{(j)}_x$ and $\theta^{(j)}_y$ are the E and N angular locations of the $j$-th micro-image in a plane tangent to the sky, while $u$ and $v$ are the corresponding Fourier axes.  The vector between a pair of telescopes on the ground corresponds to a point on the $(u,v)$ plane.  As the Earth rotates and the sky-location of object being observed changes, the corresponding point traces an ellipse in the $(u,v)$ plane.  This is illustrated in Fig.~\ref{fig:uvtracks}.  Expressions in terms of rotation matrices are well known~\cite{2020MNRAS.498.4577B}.

In intensity interferometry $V(u,v)$ is not measured directly, but inferred from non-Poissonian perturbations to photon statistics, a phenomenon known as the Hanbury~Brown and Twiss (or HBT) effect.  The HBT effect can be understood in various ways: most formally as second-order coherence in quantum optics \citep{1965RvMP...37..231M,2006RvMP...78.1267G}, but also in terms of phase-space occupancies of photons \citep{1956Natur.178.1449P}, or semi-classically as an averaging-out over incoherent fluctuations \citep{1964AmJPh..32..919M,2025Reson30.45R}.  In practical terms it provides a technique for optical interferometry over very long baselines without requiring optical coherence between the detectors, provided one can measure photon arrival times accurately.  Concretely, suppose two telescopes are each detecting photons at a rate $r$ in a given waveband and polarization state. The rate of nearly-coincident photons is not $r^2$ but
\begin{equation}
   r^2 \left(1 + |\gamma(t)|^2\right) \, |V(u,v)|^2,
\end{equation}
where $t$ is arrival-time difference, and $\gamma(t)$ is a special quantity known as the coherence function. The latter turns out to be the Fourier transform of the optical bandpass $W(\nu)$, that is,
\begin{equation}
\gamma(t) \propto \int e^{2\pi i\,\nu}\, W(\nu)\, d\nu,
\end{equation}
normalized as $\gamma(0)=1$.  The narrower the optical bandpass, the longer the time interval over which photons are correlated.

\begin{figure}
\centering
\includegraphics[width=\linewidth]{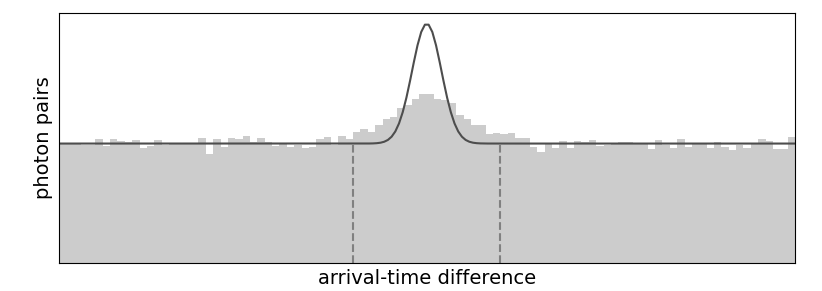}
\caption{A schematic illustration of Hanbury Brown \& Twiss correlation.  The gray histogram represents the distribution of arrival-time differences among pairs of photons.  The smooth curve represents the theoretical correlation $1+|\gamma(t)|^2$.  Timing jitter in photon detection spreads out the curve to the gray mound. The two dashed vertical lines indicate the notional time resolution.}
\label{fig:correl}
\end{figure}

Timing accuracy need not be enough to resolve $\gamma(t)$.  In practice a much higher timing jitter can be tolerated.  The effect of timing jitter is to spread out $\gamma(t)$ to an integrated quantity,
\begin{equation} \label{eq:spcoh}
   \spcoh(u,v) = |V(u,v)|^2 \, \int |\gamma(t)|^2 \, dt,
\end{equation}
known as the spatial coherence.  Fig.~\ref{fig:correl} illustrates schematically.

Let us see how to estimate $\spcoh(u,v)$ and its noise from data as in Fig.~\ref{fig:correl}.  Consider an observation of length $\tobs$. During this interval, one telescope is expected to receive $r\tobs$ photons.  For each of these photons, $r\Delta t$ photons are expected to arrive at the other telescope by chance within the interval $\Delta t$.  Meanwhile, HBT correlation will contribute a further $r\spcoh(u,v)$ photons at the second telescope. Let us now select an interval $\Delta t$ in the histogram, marked by dashed lines in the figure. The precise value of $\Delta t$ may depend on convention. Considering all pairs of photons gives,
\begin{equation}
   \textrm{Signal} = r^2 \tobs \, \spcoh(u,v).
\end{equation}
Assuming the photon statistics is nearly Poissonian gives,
\begin{equation}
   \textrm{Noise}^2 = r^2 \tobs \,
   \left(\Delta t + \spcoh(u,v)\right),
\end{equation}
In practice,
\begin{equation}
   \Delta t \gtrsim \spcoh(u,v),
\end{equation}
which gives,
\begin{equation} \label{eq:SNRv1}
   \textrm{SNR} = r\,\spcoh(u,v) \, \left(\frac{\tobs}{\Delta t}\right)^{1/2}.
\end{equation}

The photon rate $r$ depends on both the source and observing setup. To separate these dependencies, let us introduce two further quantities, as follows.
\begin{itemize}
\item Let $\Aeff$ denote the effective collecting area, meaning telescope area times optical throughput times detector efficiency.
\item Let $\Phi$ be the spectral photon flux in $\si{\textrm{photons}\;\metre^{-2}\,\second^{-1}\,\hertz^{-1}}$.
\end{itemize}
From the definition (\ref{eq:spcoh}) it follows that $\spcoh(0,0)$ is the bandwidth.  Hence,
\begin{equation} 
   r = \Aeff \, \frac{\Phi}{\spcoh(0,0)}.
\end{equation}
Furthermore, we note that distinct narrow wavebands and their two polarization states behave as independent data channels. Let us write $\Nchan$ for the number of channels.  In terms of these quantities,
\begin{equation} \label{eq:SNRv2}
  \mathrm{SNR} = \Aeff \, \left(\Nchan\,\frac{\tobs}{\Delta t}\right)^{1/2}
                 \Phi \, |V(u,v)|^2,
\end{equation}
with the telescope and instrument properties separated from the source properties.

The spectral photon flux turns out to be equivalent to an optical magnitude.  Recall the definition of AB magnitude~\citep{1983ApJ...266..713O} as a spectral energy flux of,
\begin{equation}
  f_\nu = 10^{-22.44-m_{\rm AB}/2.5} \; \si{\watt\,\metre^{-2}}.
\end{equation}
The spectral energy and photon fluxes are related by,
\begin{equation} 
   f_\nu = 2h\nu\, \Phi,
\end{equation}
(the factor of 2 appearing because $\Phi$ refers to one polarization state, whereas $f_\nu$ includes both).  Substituting we get
\begin{widetext}
\begin{equation}
  \Phi = 10^{-4.04 - m_{\rm AB}/2.5}
         \frac{\lambda}{\SI{1}{\micro\metre}} \;
      \si{\textrm{photons}\;\metre^{-2}\,\second^{-1}\,\hertz^{-1}}
\end{equation}
Returning now to the SNR expression (\ref{eq:SNRv2}) let us rewrite it as,
\begin{equation} \label{eq:SNRplausible}
  \mathrm{SNR} = \frac{\Aeff}{10^2\,\si{\metre^2}} \;
  \left( \frac{\Nchan}{10^2} \; \frac{\tobs}{10^3\,\si{\second}} \;
  \frac{10^{-11}\,\si{\second}}{\Delta t} \right)^{1/2} \;
  \frac{\Phi}{10^{-10}\,\si{\metre^{-2}\,\second^{-1}\,\hertz^{-1}}} \;
  |V(u,v)|^2.
\end{equation}
\end{widetext}
If $\Aeff$ indeed equals $10^2\,\si{\metre^2}$ and similarly for the other factors in this equation, $\textrm{SNR}=1$.  The $\Phi$ factor corresponds to $m_{\rm AB}\approx15$.  A source this faint will be observable if the other values, corresponding to the telescope and instrumentation, can be reached. In \Sref{sec:strategy}, we argue that the required instrumentation factors are plausible.

\begin{figure*}
    \centering
    \includegraphics[width=0.72\linewidth]{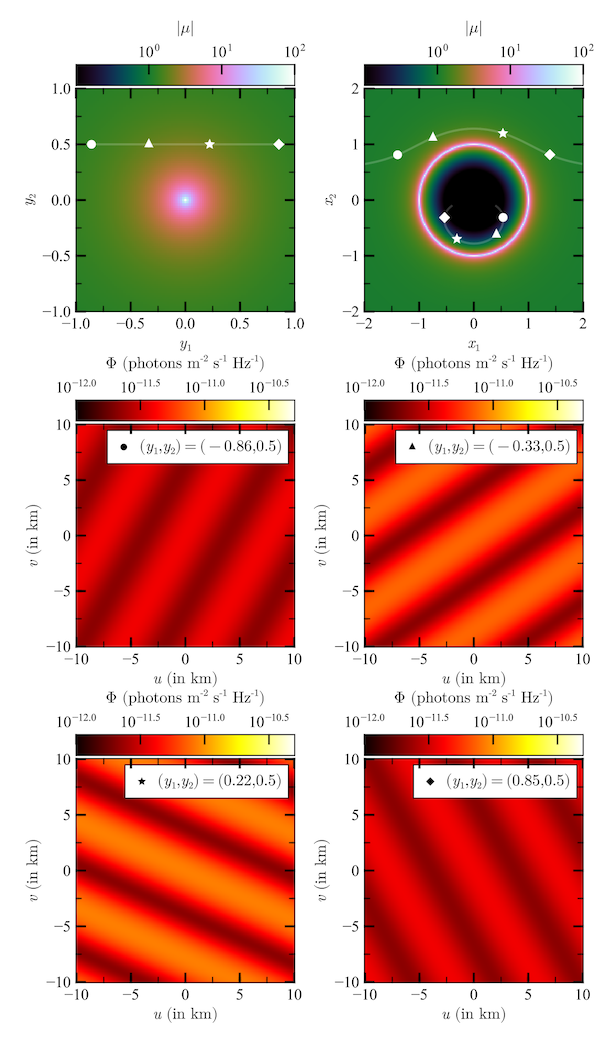}
    \caption{Photon flux for an isolated point mass lens with $M=1\,{\rm M_\odot}$. The top-left and top-right panels show the source and image plane magnification maps, respectively. The white horizontal line(s) in the top-left and top-right panels represent the mock source and the corresponding image trajectories, respectively. Four example source positions are presented by solid points in the top-left panel, and the corresponding image positions are shown in the top-right panel. The middle and bottom rows show the visibility space for these four source positions.}
    \label{fig:pml_pc}
\end{figure*}

\begin{figure*}
    \centering
    \includegraphics[width=0.72\linewidth]{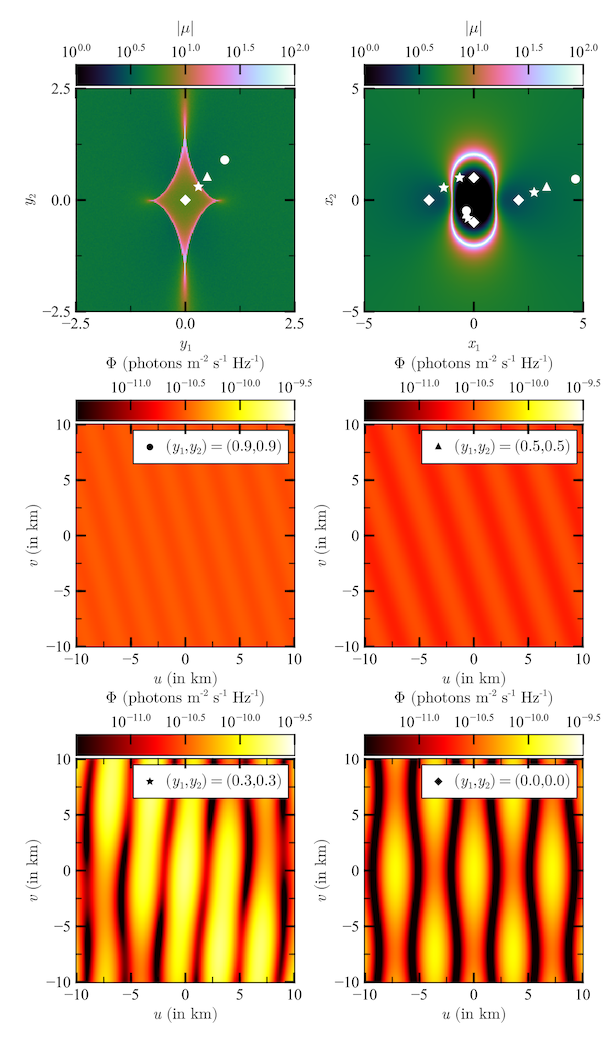}
    \caption{Photon flux for a point lens with mass~$M=1\,{\rm M_\odot}$ and external effects~$\kappa = 0.395 = \gamma$. The top-left and top-right panels show the source and image plane magnification maps. Four example source positions are presented by solid points in the top-left panel, and the corresponding image positions are shown in the top-right panel. The middle and bottom rows show the corresponding photon flux in the visibility space.}
    \label{fig:cr_pc}
\end{figure*}

\section{Point Mass Lens}
\label{sec:pml}
Gravitational lensing by isolated compact objects (such as stars, stellar remnants, and black holes) is studied using the point-mass~(i.e., Schwarzschild) lens model. The point mass lens, an axially symmetric lens model, is described by two free parameters: the point-mass lens mass ($M$) and the (angular) source position ($\beta$). The corresponding 2D projected lensing potential is given as,
\begin{equation}
    \psi(\theta) = \theta_E^2 \ln(\theta),
    \label{eq:pml_pot}
\end{equation}
where~$\theta_{\rm E}$ is the Einstein angle, a characteristic angular scale associated with the point mass lens and given as,
\begin{equation}
    \begin{split}
        \theta_{\rm E} &= \sqrt{\frac{4{\rm G}M}{\rm c^2} \frac{D_{ds}}{D_d D_s}} \\
        &= 2.''85 \times 10^{-6} \sqrt{\frac{M}{\rm M_\odot} \frac{\rm 1\,Gpc}{D_d} \frac{D_{ds}}{D_s}}
    \end{split}
    \label{eq:pml_ein}
\end{equation}
where $D_{d}$, $D_s$, and $D_{ds}$ are angular diameter distances from observer to lens, observer to source, and lens to source, respectively.

A point mass lens always leads to the  formation of two images (one minimum and one saddle-point) with separation,
\begin{equation}
    \Delta \theta = \sqrt{\beta^2 + 4\:\theta_E^2}.
    \label{eq:pml_sep}
\end{equation}
The magnifications of these two images are given as
\begin{equation}
    \mu_{\pm} = \frac{1}{2} \pm \frac{\beta^2+2\,\theta_E^2}{2\,\beta\sqrt{\beta^2+4\,\theta_E^2}},
    \label{eq:pml_mu}
\end{equation}
where `+' and `-' refer to minimum and saddle-point images, respectively. The time taken by a source to travel a distance equal to the Einstein angle, \Eref{eq:pml_ein}, is,
\begin{equation}
    \begin{split}
        t_{\rm E} &= \frac{D_s \, \theta_{\rm E}}{v_t} \\
        &= 23.70~{\rm years} \left(\frac{D_s}{\rm 1\,Gpc}\right) 
        \left(\frac{\theta_{\rm E}}{1''\times10^{-6}}\right) \left(\frac{200~{\rm km/s}}{v_t}\right),
    \end{split}
    \label{eq:tE_time}
\end{equation}
where~$v_t$ is the observed relative transverse motion between lens and source with respect to the observer~\cite{1986A&A...166...36K}. For a point mass lens of $1~{\rm M_\odot}$ and~$v_t=200$~km/s, it would take the source $\sim290$~years to move a distance equal to the Einstein angle in the source plane. Due to this relative motion, the positions and magnifications of micro-images change over time, thereby changing the observed interference pattern in the visibility space. An example of this is shown in \Fref{fig:pml_pc}. Since we have only two micro-images, the interference pattern consists of parallel bright and dark fringes perpendicular to the line joining them. When the source is relatively far from the center (i.e., the optical axis), the saddle-point image is considerably fainter compared to the minimum image, resulting in reduced fringe visibility. However, as the source moves towards higher magnification, we observe an increase in fringe visibility. In \Fref{fig:pml_pc}, we chose the source positions along the~$y_2=0.5$ line, resulting in maximum magnification of 1.6 and 0.6 for the minimum and saddle-point images, respectively. If we move the source closer to the center, both the minimum and saddle-point images will be more magnified and have similar magnifications, ultimately increasing fringe visibility.

From \Fref{fig:pml_pc}, we can infer that there are two observables, the interference fringe width and its variation with time. The first tells us about the size of the micro-image swarm~(i.e., the separation between micro-images), and the second tells us about the motion of micro-images (i.e., the relative motion of the source). In addition, as mentioned above, we note that baseline(s) of approximately ten~kilometers are sufficient to determine the fringe size from a pair of micro-images formed by a point mass lens of~$1~\rm{M_\odot}$ at cosmological distance.

\section{Point Mass lens + External Effects}
\label{sec:crl}
As mentioned above, the point mass lens model is suitable only for isolated microlenses. However, a microlens might be part of a binary system, or, in the case of strong lensing, sits inside a galaxy lens. In such cases, it is important to include external effects coming from the surroundings. Hence, a more realistic lens model is a point-mass lens with external convergence and shear\footnote{Such a lens model is also known as the Chang-Refsdal lens~\cite{1979Natur.282..561C, 1984A&A...132..168C} where source position is re-scaled to absorb the external convergence.}. Since we are primarily focused on microlensing in strongly lensed quasars, such a lens is equivalent to a point mass embedded in the lensing galaxy. The corresponding potential is given as,
\begin{equation}
    \psi(\pmb{\theta}) = \theta_E^2 \ln|\pmb{\theta}| + \frac{\kappa}{2}\left(\theta_1^2 + \theta_2^2\right) + \frac{\gamma_1}{2}\left(\theta_1^2 - \theta_2^2\right) + \gamma_2 \, \theta_1 \, \theta_2,
    \label{eq:pot_cr}
\end{equation}
where~$\kappa$ and~$(\gamma_1, \gamma_2)$ denote the convergence and external shear components at the microlens position. Here, we fix~$(\kappa, \gamma_1, \gamma_2) = (0.395, 0.395, 0)$, which is equivalent to a microlens sitting close to the minimum image. The external convergence and shear values are chosen to match the global minimum image discussed in \Sref{ssec:Huchra_min}. We have also chosen $\gamma_2=0$ such that the shear direction is along the x-axis (i.e., along the u-axis in visibility space).

For such a lens, we have a diamond caustic in the source plane, and depending on the source position (inside or outside the diamond caustic), we can have either two or four images. A pair of bright images appears or disappears when the source crosses the caustic, as shown in the top row of \Fref{fig:cr_pc}. The observed interference pattern in the visibility space for the four source positions is shown in the middle and bottom rows of \Fref{fig:cr_pc}. When the source is outside the caustic, as in a point-mass lens, we observe parallel fringes. However, when the source is inside the diamond caustic, one can observe an increase in the overall photon count and a change in the interference pattern. Looking at the bottom right panel in \Fref{fig:cr_pc}, we can see that overall fringes are aligned along the y-axis but are no longer straight. This implies that the dominant contribution comes from lensed minima images on the x-axis (i.e., along the shear direction), while saddle-point images along the y-axis are less magnified. The size of fringes along the x- and y-axis is directly related to the separation between micro-images along the same axis. Since the images along the y-axis are closer together in the sky, the required baseline is larger. Comparing the photon flux with the point mass lens discussed in the previous section, we see an overall increase in the number of photons due to the macro-magnification resulting from the external effects. In addition, the caustic also contains a finite area, increasing the overall probability for a background source to cross the caustic and give rise to a large photon flux.

\begin{figure}
    \centering
    \includegraphics[width=0.85\linewidth]{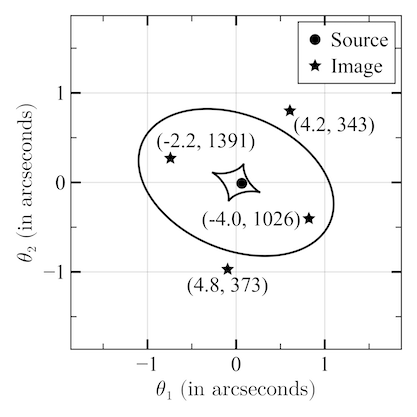}
    \caption{Quad image formation for Huchra's lens~(QSO 2237+0305). The solid curves represent the caustic and critical curves. The solid circle represents the source position, and the corresponding image positions are shown by stars. For each image position, we also show the macro-magnification and microlens surface density (in~${\rm M_\odot/pc^2}$).}
    \label{fig:huchra_sim}
\end{figure}

\begin{figure*}
    \centering
    \includegraphics[width=0.82\linewidth]{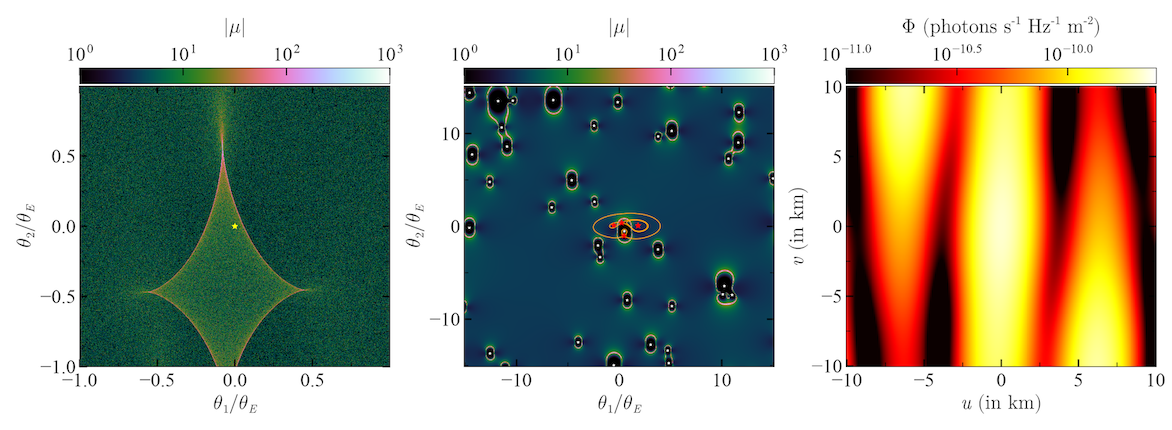}
    \includegraphics[width=0.82\linewidth]{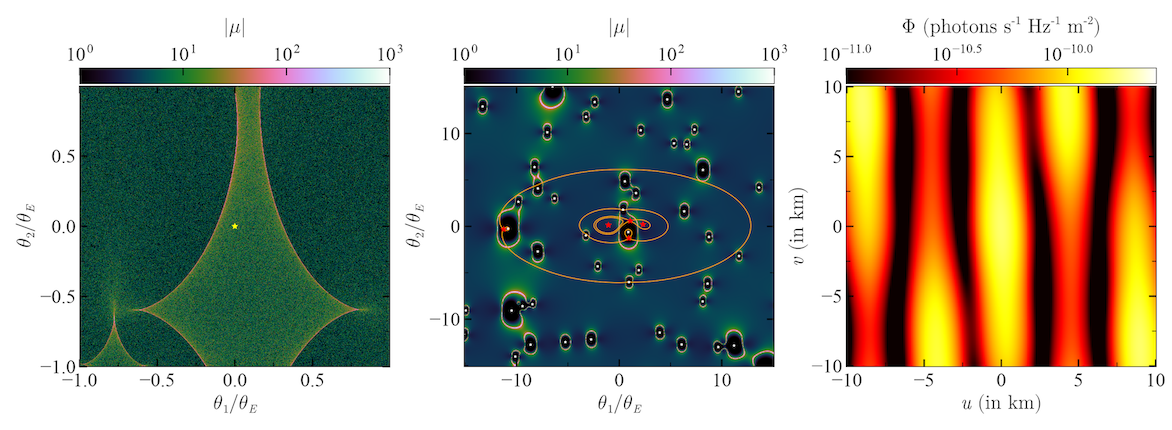}
    \includegraphics[width=0.82\linewidth]{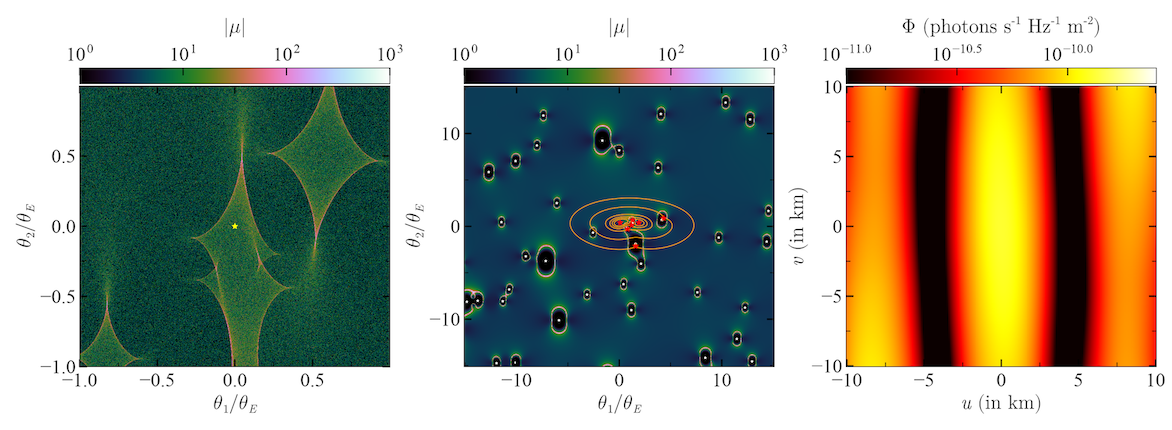}
    \includegraphics[width=0.82\linewidth]{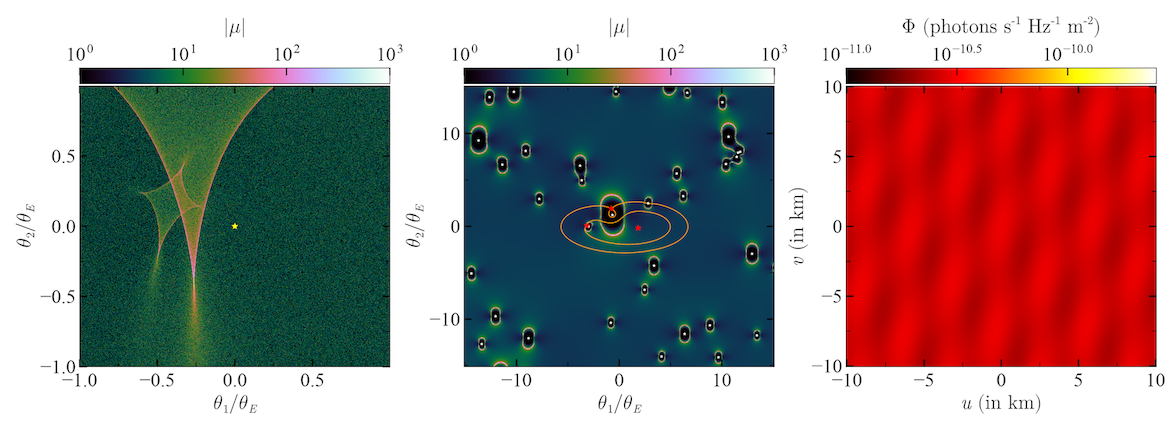}
    \caption{Microlensing and photon flux for (global) minimum image of Huchra's lens. The left column shows the source-plane magnification maps. The source position is fixed to~$(0,0)$ shown by the yellow star. The middle column shows the corresponding magnification maps of the image plane. The microlens positions are shown by white stars, whereas micro-images~(with~$|\mu|>10^{-2}$) are shown by red stars. The yellow contours represent the time delay contours corresponding to saddle points in the time-delay surface. The right column shows the corresponding visibility space maps. While constructing the visibility space maps, we consider contributions from all micro-images.}
    \label{fig:im1_pc}
\end{figure*}

\begin{figure*}
    \centering
    \includegraphics[width=0.82\linewidth]{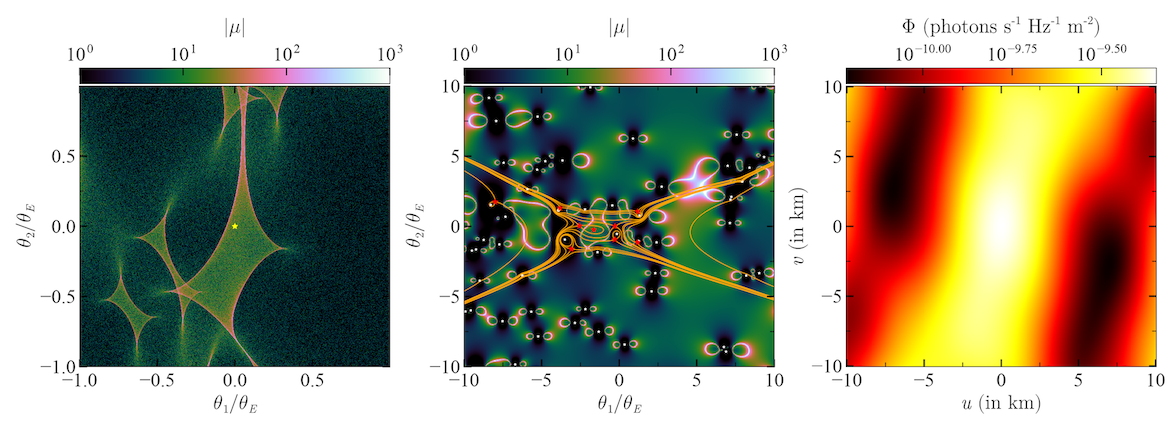}
    \includegraphics[width=0.82\linewidth]{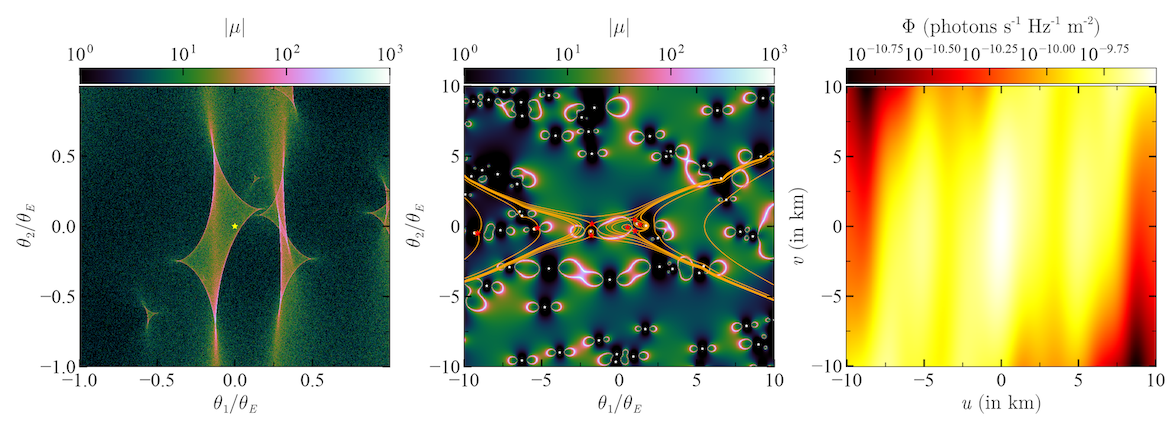}
    \includegraphics[width=0.82\linewidth]{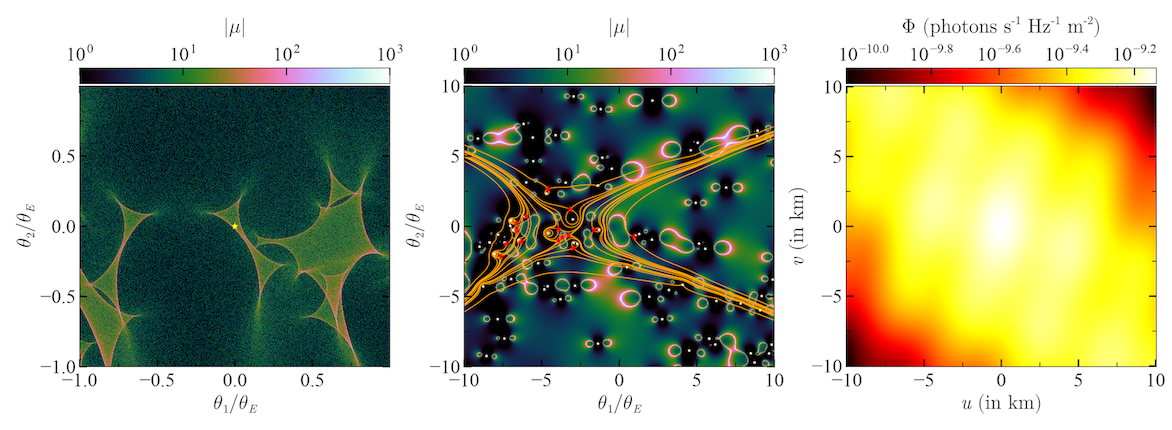}
    \includegraphics[width=0.82\linewidth]{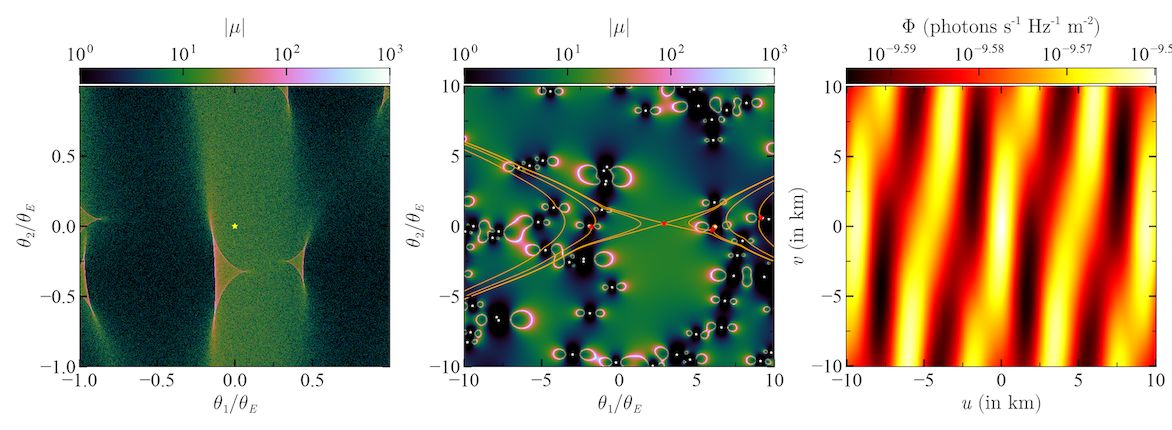}
    \caption{Microlensing and photon flux for a macro-saddle-point image of Huchra's lens. The left column shows the source-plane magnification maps. The source position is fixed to~$(0,0)$ shown by the yellow star. The middle column shows the corresponding magnification maps of the image plane. The microlens positions are shown by white stars, whereas micro-images~(with~$|\mu|>10^{-2}$) are shown by red stars. The yellow contours represent the time delay contours corresponding to saddle points in the time-delay surface. The right column shows the corresponding visibility space maps. While constructing the visibility space maps, we consider contributions from all micro-images. For the visibility plots in each panel, we adjust the color bar values to better capture the variation.}
    \label{fig:im3_pc}
\end{figure*}

\section{Huchra's lens (QSO~2237+0305)}
\label{sec:cross}
In an actual case of quasar lensing, the galaxy lens contains a large number of microlenses corresponding to stars, stellar remnants (and possible compact dark matter). In such a case, to gain insights into the observed quantities, we require an even more realistic simulation. In this section, we simulate Huchra's lens. Following~\cite{2010ApJ...719.1481V}, we use a singular isothermal ellipsoid~(SIE) lens model with velocity dispersion~($v_d$) of 182.8~km/s, ellipticity~($\epsilon$) of~0.209, and position angle of~$\ang{157.55}$~(measured with repsect to x-axis in counter-clockwise direction) to simulate a strong lens system as shown in \Fref{fig:huchra_sim}. We have the formation of four images, of which two are minima~(outside the critical curve), and two are saddle-points (inside the critical curve). 

Locally close to any of these images, to first order, the effect of the overall galaxy can be approximated by constant convergence and shear values as done in the previous section. With that, the updated lensing potential for microlensing of any of the four images is given as,
\begin{equation}
    \begin{split}
        \psi(\pmb{\theta}) = \sum_{i=1}^n \frac{M_i}{M_s} \theta_{E,s}^2\ln(|\pmb{\theta} - \pmb{\theta}_i'|) +\frac{\kappa_s}{2}\left(\theta_1^2 + \theta_2^2\right) + \\ \frac{\gamma_1}{2}\left(\theta_1^2 - \theta_2^2\right) + \gamma_2 \, \theta_1 \, \theta_2,        
    \end{split}
\end{equation}
where $M_i$ is the mass of $i$-th point mass lens situated at~$\pmb{\theta}_i'$ and $M_s$ is an arbitrary mass scale with~$\theta_{E,s}$ being the corresponding Einstein angle. $\kappa_s \equiv \kappa - \kappa_\star$ represents the smooth convergence with $\kappa$ and $\kappa_\star$ being the total convergence and convergence in the form of microlenses, respectively. $(\gamma_1, \gamma_2)$ represents the shear due to the overall galaxy lens at the image position. We again choose $\gamma_2=0$ such that the shear direction is along the x-axis (i.e., along the u-axis in visibility space). Here, we assume that all dark matter is in smooth form and microlenses only consist of stars. To calculate the stellar microlens density at each macro-image position, following~\cite{2019MNRAS.483.5583V}, we model the lens galaxy light distribution assuming a two-dimensional S\'ersic profile with an effective radius of $0.55''$ following the CASTLES-like lensed quasar population. To simulate microlens population, we use Salpeter~\cite{1955ApJ...121..161S} mass function in~$[0.08, 1.5]~{\rm M_\odot}$ range. In the following sub-section, we choose one (global) minimum and one saddle-point image to study the microlensing and its effects on the photon flux.

\subsection{Macro-minimum: $(\mu, \Sigma_\star)=(4.8, 373~{\rm M_\odot/pc^2})$}
\label{ssec:Huchra_min}
Unlike the previous section, where we only had two or four micro-images, a population of $n>1$ microlenses (without external effects) can give rise to a maximum of~$5(n-1)$ micro-images~\cite{2004math......1188K}. That said, in realistic scenarios with typical macro-magnification values, only a handful of micro-images carry most of the magnification, and one can only hope to observe (contribution from) these micro-images~\cite{2011MNRAS.411.1671S}. Some examples for the global minimum image are shown in \Fref{fig:im1_pc}. Here, in the top three rows, we deliberately choose cases where the source is inside the micro-caustic. In the middle column of \Fref{fig:im1_pc}, we only show micro-images~(red stars) that has~$|\mu|>10^{-2}$. In all three realizations, only about 6 microimages satisfy the above magnification threshold. Looking at the fringe patterns, similar to the previous section, we note the formation of ellipse-like fringes along the y-axis. This is due to the fact that the shear direction is along the x-axis. Again, the fringe size along the x- and y-axes depends on the micro-image separation along those axes, which becomes clear once we compare critical curves and fringe sizes in the top three rows with each other. 

In the bottom-most row, we chose a case in which the source is outside the micro-caustic. In such a case, nearly all of the magnification is carried by the micro-minimum image. We also observe the formation of two micro-saddles in the middle panel. However, since their magnification is much lower than micro-minimum, the fringe visibility is significantly reduced (as seen in the corresponding right panel) compared to cases where the source is inside the micro-caustic.

\subsection{Macro-saddle-point: $(\mu, \Sigma_\star)=(-4.0, 1026~{\rm M_\odot/pc^2})$}
\label{ssec:Huchra_sad}
For a macro-saddle-point image, instead of diamond micro-caustics, a microlens produces a pair of triangular caustics, as we can see in the left column in \Fref{fig:im3_pc}. The other point to note is that, although the macro-magnification is similar, the triangular caustics are relatively small in size compared to the diamond caustic for the minimum image. However, at the same time, the maximum magnification that can be achieved during micro-caustic crossing will be larger for the saddle-point. In addition, along the shear direction~(i.e., x-axis), we have twice as many crossings for the macro-saddle-point as for the macro-minimum (assuming no micro-caustic overlap), leading to more frequent variation in the observed photon flux.

In the first three rows of \Fref{fig:im3_pc}, we again selected cases where the source is inside the micro-caustic. However, in the third row, we chose an instance where the source is relatively close to the micro-caustic. Comparing with \Fref{fig:im1_pc}, we see more complex fringe patterns in the visibility space stemming from the saddle-point nature of the macro-image and higher microlens density. We note an increase in the number of micro-images with~$|\mu|\geq10^{-2}$ (represented by red stars in the middle column) as well as in the overall photon flux. Looking at the bottom-most row, we note that even when the source is outside the micro-caustic, the photon flux can exceed the macro-minimum image values, described in the previous sub-section. Although the fringe visibility is again low, we only see a $\sim10\%$ difference between maximum and minimum values of photon counts.

Due to the macro-image being a saddle-point, we observe large de-magnified regions in both the source and image planes, which are absent in macro-minimum images. This implies that for long time periods, the photon flux can remain very low. However, once the source approaches and crosses a micro-caustics, we can see sudden large spikes in photon flux.

\begin{figure}
    \centering
    \includegraphics[width=0.85\linewidth]{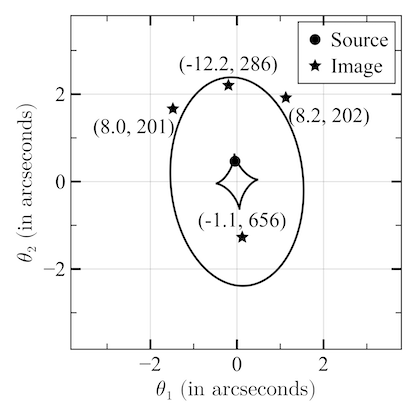}
    \caption{Quad image formation for PS~J0147+4630 lens system. The solid curves represent the caustic and critical curves. The solid circle represents the source position, and the corresponding image positions are shown by stars. For each image position, we also show the macro-magnification and microlens surface density (in~${\rm M_\odot/pc^2}$).}
    \label{fig:psj0147_sim}
\end{figure}

\begin{figure*}
    \centering
    \includegraphics[width=0.82\linewidth]{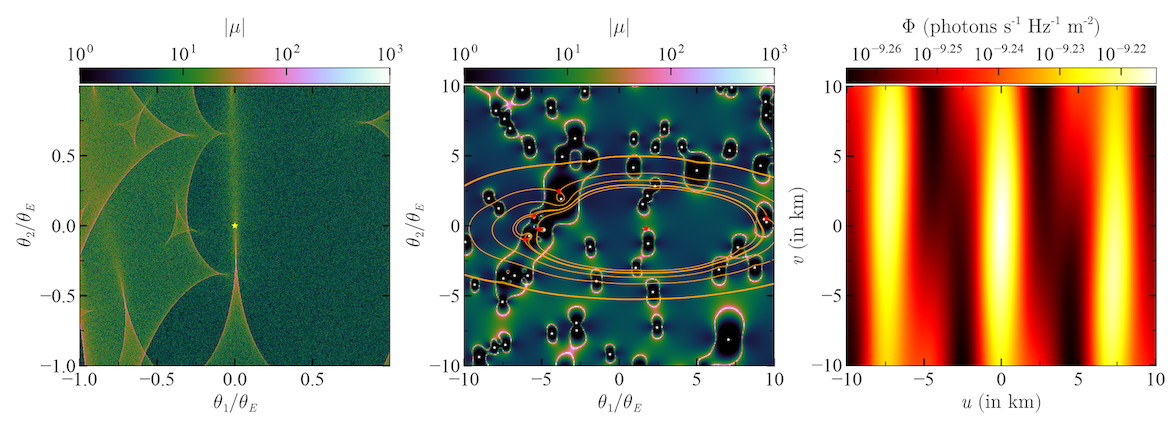}
    \includegraphics[width=0.82\linewidth]{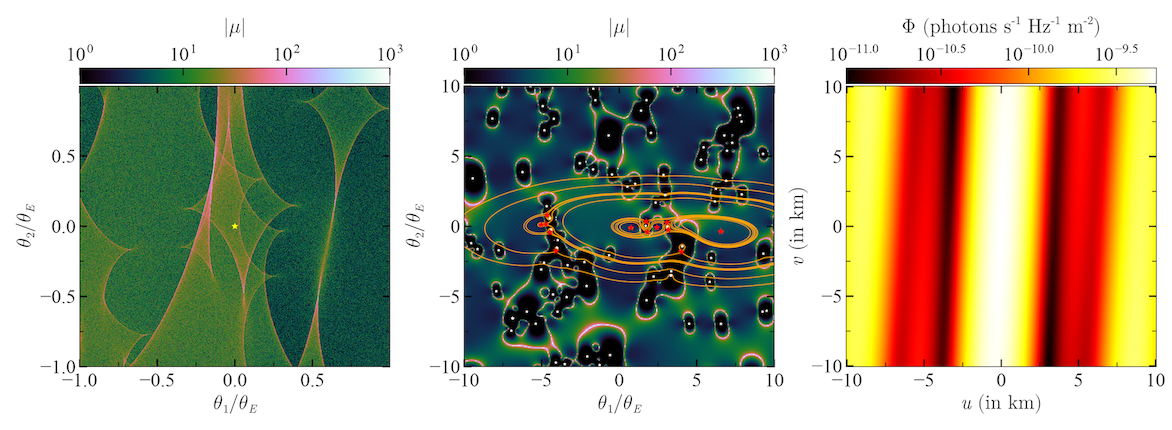}
    \includegraphics[width=0.82\linewidth]{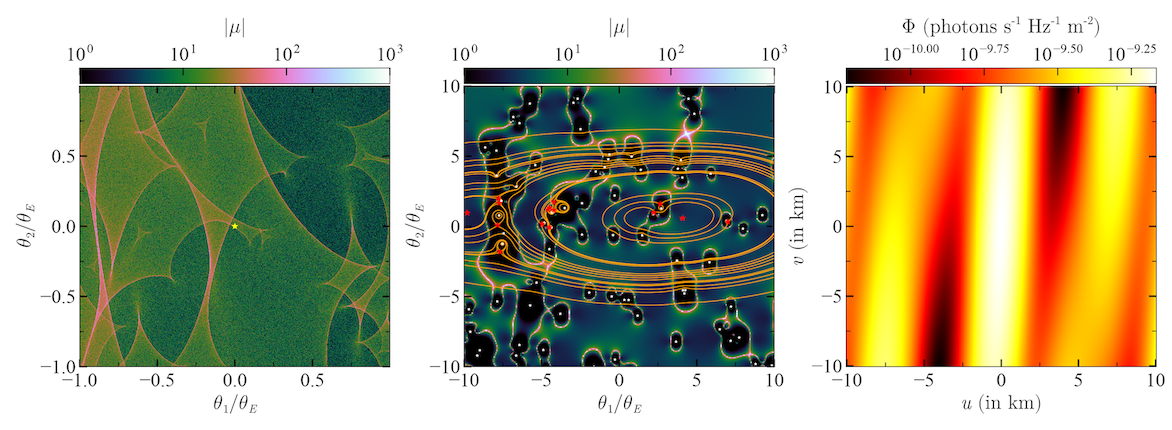}
    \includegraphics[width=0.82\linewidth]{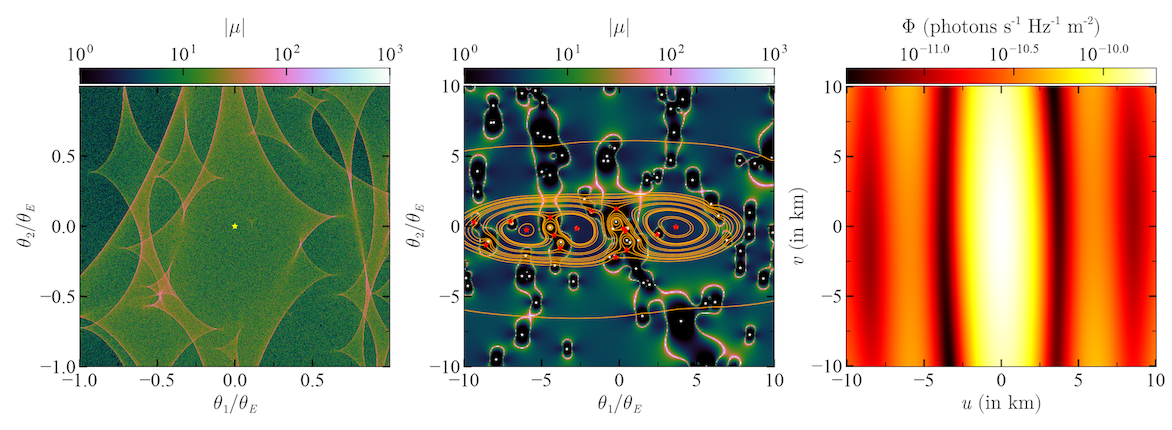}
    \caption{Microlensing and photon flux for (global) minimum image of PSJ0147 lens. The left column shows the source-plane magnification maps. The source position is fixed to~$(0,0)$ shown by the yellow star. The middle column shows the corresponding magnification maps of the image plane. The microlens positions are shown by white stars, whereas micro-images~(with~$|\mu|>10^{-2}$) are shown by red stars. The yellow contours represent the time delay contours corresponding to saddle points in the time-delay surface. The right column shows the corresponding visibility space maps. While constructing the visibility space maps, we consider contributions from all micro-images. The color bars for visibility maps are adjusted to highlight variations.}
    \label{fig:im1_pc_psj}
\end{figure*}

\begin{figure*}
    \centering
    \includegraphics[width=0.82\linewidth]{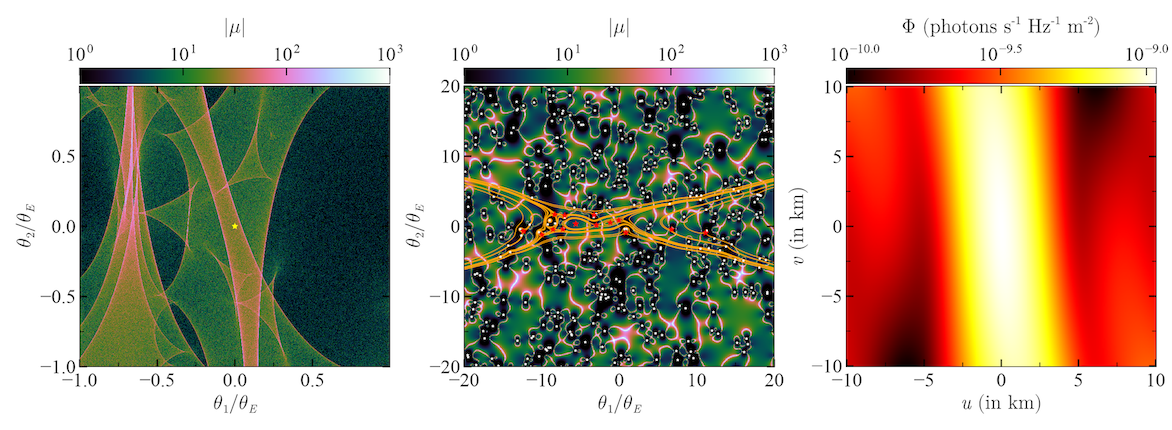}
    \includegraphics[width=0.82\linewidth]{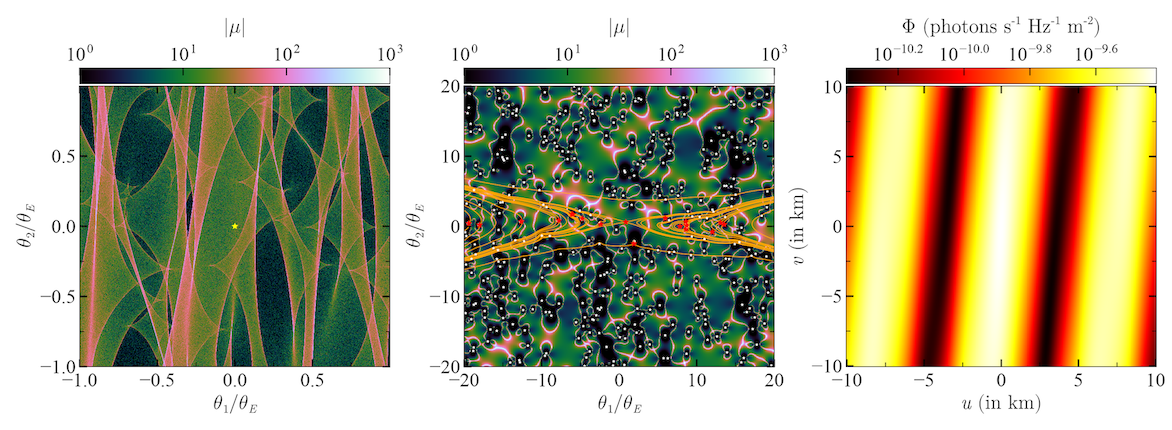}
    \includegraphics[width=0.82\linewidth]{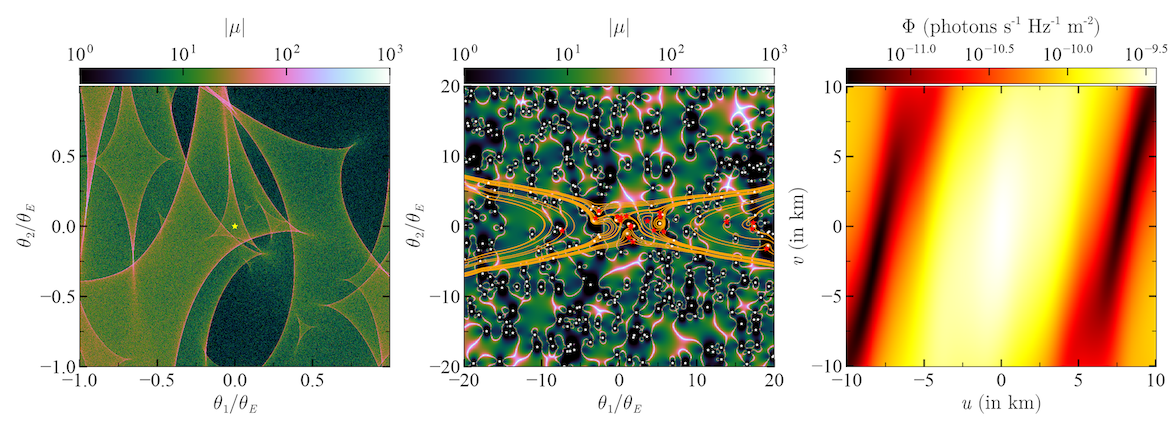}
    \includegraphics[width=0.82\linewidth]{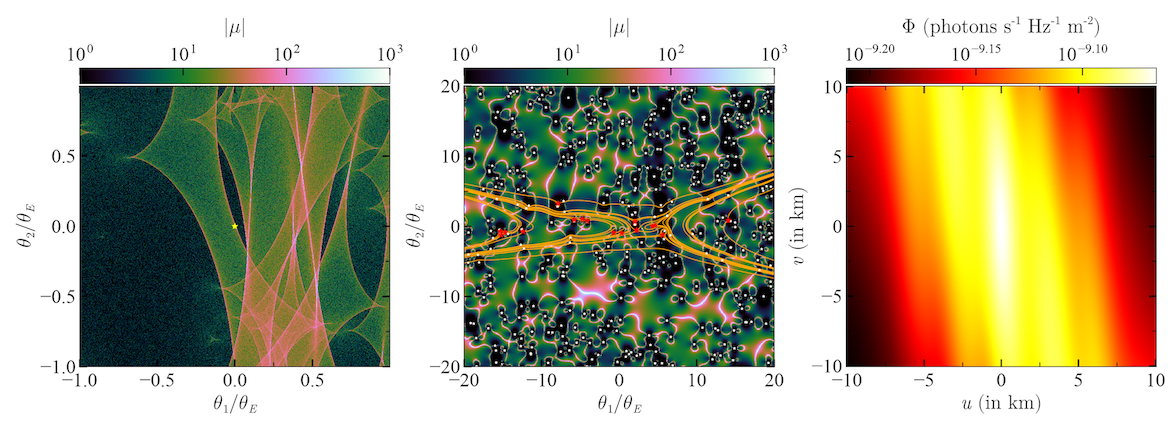}
    \caption{Microlensing and photon flux for saddle-point image of PSJ0147 lens. The left column shows the source-plane magnification maps. The source position is fixed to~$(0,0)$ shown by the yellow star. The middle column shows the corresponding magnification maps of the image plane. The microlens positions are shown by white stars, whereas micro-images~(with~$|\mu|>10^{-2}$) are shown by red stars. The yellow contours represent the time delay contours corresponding to saddle points in the time-delay surface. The right column shows the corresponding visibility space maps. While constructing the visibility space maps, we consider contributions from all micro-images. The color bars for visibility maps are adjusted to highlight variations.}
    \label{fig:im3_pc_psj}
\end{figure*}

\section{PS~J0147+4630}
\label{sec:psj0147}
Another bright lensed quasar is PS~J0147+4630 (hereafter, PSJ0147) with a quad image formation and source and lens redshifts as~$z_d=0.5716$ and~$z_s=2.341$, respectively~\cite{2017ApJ...844...90B, 2017A&A...605L...8L, 2018MNRAS.475.3086L}. For this lens redshift, the angular diameter distance to the lens is 1347.70~Kpc and~$1''=6.53$~Kpc. From \Eref{eq:tE_time}, the time taken for the source to cross the Einstein angle corresponding to~$1~{\rm M_\odot}$ is $\sim77$~years, which is $\sim$~four times smaller than Huchra's lens. The brightest image of PSJ0147 has an apparent magnitude of 15.77~AB in the G-band\footnote{\url{https://research.ast.cam.ac.uk/lensedquasars/indiv/PSJ0147+4630.html}}, roughly one magnitude brighter than Huchra's lens.

For our work, we fit a simple SIE lens model to the observed PSJ0147 image formation using \href{https://akmeena766.github.io/LensFactory.jl/}{\texttt{LensFactory.jl}}. The critical curve and caustic corresponding to the best-fit lens model, along with magnification and stellar surface density for each image, are shown in \Fref{fig:psj0147_sim}. To determine the surface density, following~\cite{2019MNRAS.483.5649S}, we again use a S\'esric profile with an effective radius of~$3.45''$. With the above, the lensing-corrected magnitude for the brightest image is 18.49~AB. The primary difference between Huchra's lens and PSJ0147 is the macro-magnification at the image positions. The macro-images of PSJ0147 have large magnification values. Similar to the previous section, we again choose one macro-minimum and one macro-saddle-point and simulate the corresponding visibility space assuming the Salpeter mass function with~$[0.08, 1.5]~{\rm M_\odot}$ for stellar microlenses at the image positions.

\subsection{Macro-minimum: $(\mu, \Sigma_\star) = (8.2, 201~{\rm M_\odot/pc^2})$}
\label{ssec:psj0417_min}

\Fref{fig:im1_pc_psj} represents examples of microlensing in the global minimum image of PSJ0147. Due to the higher macro-magnification compared to Huchra's lens, in the middle column, we note that the area covered by individual critical curves has been increased. This also led to more micro-images with $|\mu|>10^{-2}$ distributed over a larger area, as evidenced by the increase in the number of red stars and their positions. In addition, we observe the formation of more micro-minimum images than in Huchra's lens discussed in~\Sref{ssec:Huchra_min}. Since a minimum image always has~$|\mu|\geq1$, these images will have a non-zero effect on the interference pattern in the visibility space. From the source plane~(i.e., left column), we note that an increase in macro-magnification has led to more overlap in micro-caustics and more micro-caustic crossings per unit time.

In the visibility maps (i.e., the right column), we left the color bar (i.e., photon count) to vary freely from the minimum to the maximum value. This allows us to study fringes in more exotic cases. For example, in the top row, the source is just outside the cusp but in a high magnification region. In such a case, the micro-minimum image will have large magnification without creating additional bright images. In such a case, we see~$\gtrsim10\%$ variation between the brightest and darkest regions in the visibility space. On the other hand, in the second row, the source sits within multiple micro-caustics, implying multiple bright micro-images, and shows larger variations in visibility between bright and dark regions. That said, we still see that along the x-axis, we only need a baseline of ten~kilometers to determine the fringe size. However, because the micro-images are formed along the shear direction~(i.e., x-axis), baselines of more than twenty~kilometers would be required to determine the size perpendicular to the shear direction (i.e., along the y-axis).

\subsection{Macro-saddle-point: $(\mu, \Sigma_\star) = (-12.2, 286~{\rm M_\odot/pc^2})$}
\label{ssec:psj0417_sad}
\Fref{fig:im3_pc_psj} represents the examples of microlensing of the bright saddle-point image of PSJ0147. Since both macro-magnification and microlens density are higher in this image than the global minimum, the corresponding micro-critical curves in the middle column exhibit corrugation, i.e., the critical curves for individual microlenses merge. As a result, the corresponding micro-caustics also exhibit corrugation as shown in the left column. This results in the formation of a diamond caustic, which we typically see for a minimum image. Due to the larger macro-magnification, we again observe an increase in the formation of micro-images at greater distances from the center. For such a macro-image, we expect to see more frequent variation in the visibility space, as the source will cross micro-caustics more often. Similar to the previous section, we observe large de-magnified regions in the source plane, implying larger variations in the observed photon flux.

\section{Observation feasibility}
\label{sec:strategy}
The current II observing campaigns with the Cherenkov telescope are targeting bright stars with apparent magnitudes~$\lesssim5$~AB in the sky. Since these facilities were primarily designed for gamma-ray astronomy, their mirror quality is inferior to that of optical telescopes. Hence, the corresponding point spread function~(PSF) is of the order of milli-radians, which limits us from going deeper than~$\sim8-10$~AB due to the night sky background~\cite{Saha:2025uG}. The (current) brightest (unlensed) quasar~(3C~273) in the sky has an apparent magnitude of~$\sim12.9$~AB at optical wavelengths. Hence, both unlensed and lensed quasars are beyond the above observing capabilities, and one needs to shift to optical telescopes to have any chance of observing them. Recently, II has also been done 1-meter class optical telescopes~\cite{2018MNRAS.480..245G, 2020MNRAS.494..218R, 2023AJ....165..117M} targeting very bright stars within our own Galaxy. Although not possible with the above 1-meter class telescopes, here, we argue that, with ongoing technical advancements, it is plausible to study micro-image swarms in the brightest lensed quasars. Looking at \Eref{eq:SNRplausible}, the various factors that affect the SNR are:
\begin{itemize}
    \item Spectral photon flux~($\Phi$): For a $16$~AB source at $\SI{0.5}{\micro\metre}$, the unlensed photon flux is~$\Phi\simeq 10^{-11}$. However, as we have seen above, during micro-caustic crossing, the observed flux can be enhanced. The enhancement factor would depend on the size of the source~(the smaller the source, the higher the magnification). Micro-caustic crossings can give rise to magnifications of~$\sim50$ in lensed quasar images. As we have seen in \Sref{sec:cross} and \Sref{sec:psj0147}, even when the source is relatively far from the micro-caustic, the photon flux can be~$\Phi>10^{-10}$. During micro-caustic crossing, assuming~$|\mu|\sim50$, the observed photon would be~$\Phi\sim 10^{-9}$. We also note that chances of getting larger magnifications are higher for macro-saddle-point images~\cite{2025A&A...701A..35N}.

    \item Effective collecting area~($A_{\rm eff}$): As the SNR is proportional to~$A_{\rm eff}$, the collecting area should be as large as possible. Considering the VLT~(8.2~m) and ELT~(39.3~m) pair as our baseline, the geometric mean of the corresponding collecting areas is~$253.1~{\rm m^2}$. However, one also needs to account for various losses, such as telescope throughput, optical system transmission, and the gap between detector pixels. Assuming that these effects decrease the effective collecting area by a factor of one- third~\cite{Walter:2023yF}, we get~$A_{\rm eff}\simeq84~{\rm m^2}$ for the VLT-ELT-like baseline.

    \item Timing resolution~($\Delta t$): As mentioned in \Sref{sec:basic_II}, uncertainty in time resolution would decrease the correlation significance. Hence, the photon detector should time-tag every photon with extreme precision. Ideally, the detector timing resolution should be of the order of atmospheric jitter of~$\sim1-10~{\rm ps}$~\cite{1974iiaa.book.....B}. With the state-of-the-art single-photon avalanche diode~(SPAD) detectors, timing uncertainty $\sim 10$~ps can be achieved~\cite{Gramuglia:2022oii, Walter:2023yF} with quantum efficiency~$\sim0.4$. Even better timing resolution of~$<10$~ps with quantum efficiency~$\gtrsim0.9$ has been achieved with superconducting nano-wire single-photon detector~(SNSPD) in laboratory \cite{2020NaPho..14..250K, Reddy:20, 2024NatCo..15.3973C}. Hence, it would be reasonable to assume that in future, we can reach a time resolution of 10~ps.

    \item Number of channels~($N_{\rm ch}$): Traditionally, II observations have mainly used one frequency channel per telescope, and only very recently people have started considering multi-wavelength channels~\cite{2024MNRAS.529.4387A}. For a baseline such as VLT-ELT, by increasing the number of channels, we can increase the coherence time. Although the SNR in individual channels will be low, the total SNR will improve as~$\sqrt{N_{\rm ch}}$.   
\end{itemize}

With the above considerations, assuming~$A_{\rm eff} = 10^2~{\rm m^2}$, $\Phi = 10^{-9}$, $\Delta t = 10~{\rm ps}$, and~$N_{\rm ch}=10^3$ seems feasible in the future during micro-caustic crossing(s). This would lead to SNR~$\simeq30$ for~$T=10^3$~s. With the above idealized scenario, we have a factor of five-to-six to play around to consider a more feasible observing setup. For example, if we decrease the timing resolution and number of channels each by factors of five, we can still get an SNR~$\simeq5$. Instead, we could drop the effective telescope area~($A_{\rm eff}$) or the photon flux~($\Phi$) during the micro-caustic crossing by a factor of five and still get~$\simeq5$. It seems more plausible to consider variation in all of the above factors together to come up with the most practical instrument setup to make such observations a reality.

In II observation(s) of lensed quasars, observations at optical (and other wavelengths) are expected to play a crucial role. With optical observations, we can build very robust lens models that would allow us to determine the macro-magnification as well as the shear direction. Since fringes tend to be perpendicular to the shear direction, we can determine in advance whether the $u,v$ coverage is optimal by ensuring that we are likely to cover the full fringe to estimate its size. In addition, since micro-caustic crossings occur randomly, rather than targeting the source for long periods with II, we can rely on optical follow-up to determine the optimal window for II observations. With optical surface brightness models, we can model the contribution from stellar microlenses (given an initial mass function) at the image positions, predict the size of the micro-image swarm, and compare it with the II observations. In addition, including kinematic observations would also help in breaking lensing degeneracies.

\section{Conclusions}
\label{sec:conclusion}
Each strongly lensed quasar image is composed of a swarm of micro-images separated by micro-arcseconds, due to the presence of microlenses within the strong lens. The number of micro-images, their properties, and the swarm angular size depend on the underlying microlens mass function and the macro-magnification. Hence, determining the micro-image swarm properties is crucial for constraining the stellar microlens mass function~(as well as the compact dark matter fraction) in distant lensing galaxies. Since the required resolution is of the order of micro-arcseconds, neither current nor (planned) future optical telescope would be able to resolve such micro-image swarms. One (and maybe only) way forward is to target the brightest lensed quasar image with intensity interferometry~(II).

In our current work, we study the prospects of brightest lensed quasar observations with II and the feasibility of determining the micro-image swarm properties. We consider QSO~2237+0305 and PS~J0147+4630, two of the brightest lensed quasars in the sky, and show that, with the planned facilities, it is plausible to measure the size of the micro-image swarm. We find that for both macro-minimum and macro-saddle-point of QSO~2237+0305, only~$\lesssim6$ micro-images have~$|\mu|\gtrsim10^{-2}$ when the source is inside a micro-caustic. On the other hand, for PS~J0147+4630, due to larger macro-magnification values, the number of micro-images with~$|\mu|\gtrsim10^{-2}$ as well as the extent of micro-images along the shear axis increases. In all simulated cases, we find that a baseline of 20~km is sufficient to determine the size of the micro-image swarm parallel to the shear directions. However, the same is not always true for perpendicular to the shear direction, depending on the values of tangential and radial macro-magnifications. Although not possible with current facilities, considering the ongoing advances in detector technology and the upcoming large optical telescope, we show that such observations are plausible for the above-mentioned lensed quasars.

Apart from the technological challenges, another limitation of such an observing campaign is the number of sources in the sky that could provide sufficient SNR. In our current work, we have only chosen two of the brightest quasars. Other possible candidates with similar brightness are B1422+231\footnote{\url{https://research.ast.cam.ac.uk/lensedquasars/indiv/B1422+231.html}}, HS~0810+2554\footnote{\url{https://research.ast.cam.ac.uk/lensedquasars/indiv/HS0810+2554.html}} and J1330-0905~\cite{2026arXiv260413152D}. These seem to be the only suitable targets for II in the foreseeable future. II is photon-starved. Hence, the lensed quasar macro-images need to be intrinsically bright.  With too few targets, it might be that whatever we infer about the microlenses in the lens galaxy (or the source itself) gives us only a biased view. Nevertheless, pursuing these targets with II remains worthwhile, as it offers a complementary way to probe microlensing at the extragalactic scale.

\section*{Acknowledgments}
The authors thank Luke Weisenbach and Liliya Williams for their useful comments. AKM acknowledges support from the Start-up Grant IE/CARE-25-0305 provided by the IISc, Bengaluru, India.  This research has made use of NASA's Astrophysics Data System Bibliographic Services.

The work utilizes the following software packages:
\href{https://julialang.org/}{\texttt{Julia}}~\cite{2014arXiv1411.1607B},
\href{https://akmeena766.github.io/LensFactory.jl/}{\texttt{LensFactory.jl}}.

\section*{Data Availability}
The data supporting the findings of this article can be generated using the code available at: \url{https://github.com/akmeena766/GL_II}.

\def\mnras{MNRAS}
\def\aap{A\&A}
\def\aj{AJ}
\def\apj{ApJ}
\def\apjl{ApJL}
\def\apjs{ApJS}
\def\nar{Nature Ast}
\textit{}
\bibliographystyle{apsrev4-2}
\bibliography{Reference} 

@ARTICLE{1921ApJ....53..249M,
       author = {{Michelson}, A.~A. and {Pease}, F.~G.},
        title = "{Measurement of the Diameter of {\ensuremath{\alpha}} Orionis with the Interferometer.}",
      journal = {\apj},
         year = 1921,
        month = may,
       volume = {53},
        pages = {249-259},
          doi = {10.1086/142603},
       adsurl = {https://ui.adsabs.harvard.edu/abs/1921ApJ....53..249M}
}

@ARTICLE{1952Natur.170.1061H,
       author = {{Hanbury Brown}, R. and {Jennison}, R.~C. and {Gupta}, M.~K. Das},
        title = "{Apparent Angular Sizes of Discrete Radio Sources: Observations at Jodrell Bank, Manchester}",
      journal = {\nat},
         year = 1952,
        month = dec,
       volume = {170},
       number = {4338},
        pages = {1061-1063},
          doi = {10.1038/1701061a0},
       adsurl = {https://ui.adsabs.harvard.edu/abs/1952Natur.170.1061H}
}

@ARTICLE{1955ApJ...121..161S,
       author = {{Salpeter}, Edwin E.},
        title = "{The Luminosity Function and Stellar Evolution.}",
      journal = {\apj},
         year = 1955,
        month = jan,
       volume = {121},
        pages = {161},
          doi = {10.1086/145971},
       adsurl = {https://ui.adsabs.harvard.edu/abs/1955ApJ...121..161S}
}

@ARTICLE{1956Natur.178.1449P,
       author = {{Purcell}, E.~M.},
        title = "{The Question of Correlation between Photons in Coherent Light Rays}",
      journal = {\nat},
         year = 1956,
        month = dec,
       volume = {178},
       number = {4548},
        pages = {1449-1450},
          doi = {10.1038/1781449a0},
       adsurl = {https://ui.adsabs.harvard.edu/abs/1956Natur.178.1449P}
}

@ARTICLE{1964AmJPh..32..919M,
       author = {{Martienssen}, W. and {Spiller}, E.},
        title = "{Coherence and Fluctuations in Light Beams}",
      journal = {American Journal of Physics},
         year = 1964,
        month = dec,
       volume = {32},
       number = {12},
        pages = {919-926},
          doi = {10.1119/1.1970023},
       adsurl = {https://ui.adsabs.harvard.edu/abs/1964AmJPh..32..919M}
}

@ARTICLE{1965RvMP...37..231M,
   author = {{Mandel}, L. and {Wolf}, E.},
    title = "{Coherence Properties of Optical Fields}",
  journal = {Reviews of Modern Physics},
     year = 1965,
    month = apr,
   volume = 37,
    pages = {231-287},
      doi = {10.1103/RevModPhys.37.231},
   adsurl = {https://ui.adsabs.harvard.edu/abs/1965RvMP...37..231M}
}

@ARTICLE{1974MNRAS.167..121H,
       author = {{Hanbury Brown}, R. and {Davis}, J. and {Allen}, L.~R.},
        title = "{The Angular Diameters of 32 Stars}",
      journal = {\mnras},
         year = 1974,
        month = apr,
       volume = {167},
        pages = {121-136},
          doi = {10.1093/mnras/167.1.121},
       adsurl = {https://ui.adsabs.harvard.edu/abs/1974MNRAS.167..121H}
}

@BOOK{1974iiaa.book.....B,
       author = {{Brown}, R.~H.},
        title = "{The intensity interferometer; its application to astronomy}",
         year = 1974,
       adsurl = {https://ui.adsabs.harvard.edu/abs/1974iiaa.book.....B}
}

@ARTICLE{1979Natur.282..561C,
       author = {{Chang}, K. and {Refsdal}, S.},
        title = "{Flux variations of QSO 0957 + 561 A, B and image splitting by stars near the light path}",
      journal = {\nat},
         year = 1979,
        month = dec,
       volume = {282},
       number = {5739},
        pages = {561-564},
          doi = {10.1038/282561a0},
       adsurl = {https://ui.adsabs.harvard.edu/abs/1979Natur.282..561C}
}

@ARTICLE{1983ApJ...266..713O,
       author = {{Oke}, J.~B. and {Gunn}, J.~E.},
        title = "{Secondary standard stars for absolute spectrophotometry.}",
      journal = {\apj},
         year = 1983,
        month = mar,
       volume = {266},
        pages = {713-717},
          doi = {10.1086/160817},
       adsurl = {https://ui.adsabs.harvard.edu/abs/1983ApJ...266..713O}
}

@ARTICLE{1984A&A...132..168C,
       author = {{Chang}, K. and {Refsdal}, S.},
        title = "{Star disturbances in gravitational lens galaxies.}",
      journal = {\aap},
         year = 1984,
        month = mar,
       volume = {132},
        pages = {168-178},
       adsurl = {https://ui.adsabs.harvard.edu/abs/1984A&A...132..168C}
}

@ARTICLE{1986A&A...166...36K,
       author = {{Kayser}, R. and {Refsdal}, S. and {Stabell}, R.},
        title = "{Astrophysical applications of gravitational micro-lensing.}",
      journal = {\aap},
         year = 1986,
        month = sep,
       volume = {166},
        pages = {36-52},
       adsurl = {https://ui.adsabs.harvard.edu/abs/1986A&A...166...36K}
}

@BOOK{1992grle.book.....S,
       author = {{Schneider}, Peter and {Ehlers}, J{\"u}rgen and {Falco}, Emilio E.},
        title = "{Gravitational Lenses}",
         year = 1992,
          doi = {10.1007/978-3-662-03758-4},
       adsurl = {https://ui.adsabs.harvard.edu/abs/1992grle.book.....S}
}

@ARTICLE{1996astro.ph..6001N,
       author = {{Narayan}, Ramesh and {Bartelmann}, Matthias},
        title = "{Lectures on Gravitational Lensing}",
      journal = {arXiv e-prints},
         year = 1996,
        month = jun,
          eid = {astro-ph/9606001},
        pages = {astro-ph/9606001},
          doi = {10.48550/arXiv.astro-ph/9606001},
archivePrefix = {arXiv},
       eprint = {astro-ph/9606001},
 primaryClass = {astro-ph},
       adsurl = {https://ui.adsabs.harvard.edu/abs/1996astro.ph..6001N}
}

@ARTICLE{2004math......1188K,
       author = {{Khavinson}, Dmitry and {Neumann}, Genevra},
        title = "{On the number of zeros of certain rational harmonic functions}",
      journal = {arXiv Mathematics e-prints},
         year = 2004,
        month = jan,
          eid = {math/0401188},
        pages = {math/0401188},
          doi = {10.48550/arXiv.math/0401188},
archivePrefix = {arXiv},
       eprint = {math/0401188},
 primaryClass = {math.CV},
       adsurl = {https://ui.adsabs.harvard.edu/abs/2004math......1188K}
}

@ARTICLE{2006RvMP...78.1267G,
       author = {{Glauber}, Roy J.},
        title = "{Nobel Lecture: One hundred years of light quanta}",
      journal = {Reviews of Modern Physics},
         year = 2006,
        month = oct,
       volume = {78},
       number = {4},
        pages = {1267-1278},
          doi = {10.1103/RevModPhys.78.1267},
       adsurl = {https://ui.adsabs.harvard.edu/abs/2006RvMP...78.1267G}
}

@ARTICLE{2007A&A...469..387T,
       author = {{Tisserand}, P. and {Le Guillou}, L. and {Afonso}, C. and {Albert}, J.~N. and {Andersen}, J. and {Ansari}, R. and {Aubourg}, {\'E}. and {Bareyre}, P. and {Beaulieu}, J.~P. and {Charlot}, X. and {Coutures}, C. and {Ferlet}, R. and {Fouqu{\'e}}, P. and {Glicenstein}, J.~F. and {Goldman}, B. and {Gould}, A. and {Graff}, D. and {Gros}, M. and {Haissinski}, J. and {Hamadache}, C. and {de Kat}, J. and {Lasserre}, T. and {Lesquoy}, {\'E}. and {Loup}, C. and {Magneville}, C. and {Marquette}, J.~B. and {Maurice}, {\'E}. and {Maury}, A. and {Milsztajn}, A. and {Moniez}, M. and {Palanque-Delabrouille}, N. and {Perdereau}, O. and {Rahal}, Y.~R. and {Rich}, J. and {Spiro}, M. and {Vidal-Madjar}, A. and {Vigroux}, L. and {Zylberajch}, S. and {EROS-2 Collaboration}},
        title = "{Limits on the Macho content of the Galactic Halo from the EROS-2 Survey of the Magellanic Clouds}",
      journal = {\aap},
         year = 2007,
        month = jul,
       volume = {469},
       number = {2},
        pages = {387-404},
          doi = {10.1051/0004-6361:20066017},
archivePrefix = {arXiv},
       eprint = {astro-ph/0607207},
 primaryClass = {astro-ph},
       adsurl = {https://ui.adsabs.harvard.edu/abs/2007A&A...469..387T}
}

@ARTICLE{2010ApJ...719.1481V,
       author = {{van de Ven}, Glenn and {Falc{\'o}n-Barroso}, Jes{\'u}s and {McDermid}, Richard M. and {Cappellari}, Michele and {Miller}, Bryan W. and {de Zeeuw}, P. Tim},
        title = "{The Einstein Cross: Constraint on Dark Matter from Stellar Dynamics and Gravitational Lensing}",
      journal = {\apj},
         year = 2010,
        month = aug,
       volume = {719},
       number = {2},
        pages = {1481-1496},
          doi = {10.1088/0004-637X/719/2/1481},
archivePrefix = {arXiv},
       eprint = {0807.4175},
 primaryClass = {astro-ph},
       adsurl = {https://ui.adsabs.harvard.edu/abs/2010ApJ...719.1481V}
}

@ARTICLE{2011MNRAS.411.1671S,
       author = {{Saha}, Prasenjit and {Williams}, Liliya L.~R.},
        title = "{Understanding micro-image configurations in quasar microlensing}",
      journal = {\mnras},
         year = 2011,
        month = mar,
       volume = {411},
       number = {3},
        pages = {1671-1677},
          doi = {10.1111/j.1365-2966.2010.17797.x},
archivePrefix = {arXiv},
       eprint = {1010.0006},
 primaryClass = {astro-ph.CO},
       adsurl = {https://ui.adsabs.harvard.edu/abs/2011MNRAS.411.1671S}
}

@ARTICLE{2012ApJ...746..101B,
       author = {{Boyajian}, Tabetha S. and {McAlister}, Harold A. and {van Belle}, Gerard and {Gies}, Douglas R. and {ten Brummelaar}, Theo A. and {von Braun}, Kaspar and {Farrington}, Chris and {Goldfinger}, P.~J. and {O'Brien}, David and {Parks}, J. Robert and {Richardson}, Noel D. and {Ridgway}, Stephen and {Schaefer}, Gail and {Sturmann}, Laszlo and {Sturmann}, Judit and {Touhami}, Yamina and {Turner}, Nils H. and {White}, Russel},
        title = "{Stellar Diameters and Temperatures. I. Main-sequence A, F, and G Stars}",
      journal = {\apj},
         year = 2012,
        month = feb,
       volume = {746},
       number = {1},
          eid = {101},
        pages = {101},
          doi = {10.1088/0004-637X/746/1/101},
archivePrefix = {arXiv},
       eprint = {1112.3316},
 primaryClass = {astro-ph.SR},
       adsurl = {https://ui.adsabs.harvard.edu/abs/2012ApJ...746..101B}
}

@ARTICLE{2012ApJ...757..112B,
       author = {{Boyajian}, Tabetha S. and {von Braun}, Kaspar and {van Belle}, Gerard and {McAlister}, Harold A. and {ten Brummelaar}, Theo A. and {Kane}, Stephen R. and {Muirhead}, Philip S. and {Jones}, Jeremy and {White}, Russel and {Schaefer}, Gail and {Ciardi}, David and {Henry}, Todd and {L{\'o}pez-Morales}, Mercedes and {Ridgway}, Stephen and {Gies}, Douglas and {Jao}, Wei-Chun and {Rojas-Ayala}, B{\'a}rbara and {Parks}, J. Robert and {Sturmann}, Laszlo and {Sturmann}, Judit and {Turner}, Nils H. and {Farrington}, Chris and {Goldfinger}, P.~J. and {Berger}, David H.},
        title = "{Stellar Diameters and Temperatures. II. Main-sequence K- and M-stars}",
      journal = {\apj},
         year = 2012,
        month = oct,
       volume = {757},
       number = {2},
          eid = {112},
        pages = {112},
          doi = {10.1088/0004-637X/757/2/112},
archivePrefix = {arXiv},
       eprint = {1208.2431},
 primaryClass = {astro-ph.SR},
       adsurl = {https://ui.adsabs.harvard.edu/abs/2012ApJ...757..112B}
}

@ARTICLE{2012NewAR..56..143D,
       author = {{Dravins}, Dainis and {LeBohec}, Stephan and {Jensen}, Hannes and {Nu{\~n}ez}, Paul D.},
        title = "{Stellar intensity interferometry: Prospects for sub-milliarcsecond optical imaging}",
      journal = {\nar},
         year = 2012,
        month = nov,
       volume = {56},
       number = {5},
        pages = {143-167},
          doi = {10.1016/j.newar.2012.06.001},
archivePrefix = {arXiv},
       eprint = {1207.0808},
 primaryClass = {astro-ph.IM},
       adsurl = {https://ui.adsabs.harvard.edu/abs/2012NewAR..56..143D}
}

@ARTICLE{2014arXiv1411.1607B,
       author = {{Bezanson}, Jeff and {Edelman}, Alan and {Karpinski}, Stefan and {Shah}, Viral B.},
        title = "{Julia: A Fresh Approach to Numerical Computing}",
      journal = {arXiv e-prints},
         year = 2014,
        month = nov,
          eid = {arXiv:1411.1607},
        pages = {arXiv:1411.1607},
          doi = {10.48550/arXiv.1411.1607},
archivePrefix = {arXiv},
       eprint = {1411.1607},
 primaryClass = {cs.MS},
       adsurl = {https://ui.adsabs.harvard.edu/abs/2014arXiv1411.1607B}
}

@ARTICLE{2015MNRAS.447..846B,
       author = {{Boyajian}, Tabetha and {von Braun}, Kaspar and {Feiden}, Gregory A. and {Huber}, Daniel and {Basu}, Sarbani and {Demarque}, Pierre and {Fischer}, Debra A. and {Schaefer}, Gail and {Mann}, Andrew W. and {White}, Timothy R. and {Maestro}, Vicente and {Brewer}, John and {Lamell}, C. Brooke and {Spada}, Federico and {L{\'o}pez-Morales}, Mercedes and {Ireland}, Michael and {Farrington}, Chris and {van Belle}, Gerard T. and {Kane}, Stephen R. and {Jones}, Jeremy and {ten Brummelaar}, Theo A. and {Ciardi}, David R. and {McAlister}, Harold A. and {Ridgway}, Stephen and {Goldfinger}, P.~J. and {Turner}, Nils H. and {Sturmann}, Laszlo},
        title = "{Stellar diameters and temperatures - VI. High angular resolution measurements of the transiting exoplanet host stars HD 189733 and HD 209458 and implications for models of cool dwarfs}",
      journal = {\mnras},
         year = 2015,
        month = feb,
       volume = {447},
       number = {1},
        pages = {846-857},
          doi = {10.1093/mnras/stu2502},
archivePrefix = {arXiv},
       eprint = {1411.5638},
 primaryClass = {astro-ph.SR},
       adsurl = {https://ui.adsabs.harvard.edu/abs/2015MNRAS.447..846B}
}

@ARTICLE{2016Natur.533..217R,
       author = {{Roettenbacher}, R.~M. and {Monnier}, J.~D. and {Korhonen}, H. and {Aarnio}, A.~N. and {Baron}, F. and {Che}, X. and {Harmon}, R.~O. and {K{\H{o}}v{\'a}ri}, Zs. and {Kraus}, S. and {Schaefer}, G.~H. and {Torres}, G. and {Zhao}, M. and {Ten Brummelaar}, T.~A. and {Sturmann}, J. and {Sturmann}, L.},
        title = "{No Sun-like dynamo on the active star {\ensuremath{\zeta}} Andromedae from starspot asymmetry}",
      journal = {\nat},
         year = 2016,
        month = may,
       volume = {533},
       number = {7602},
        pages = {217-220},
          doi = {10.1038/nature17444},
archivePrefix = {arXiv},
       eprint = {1709.10107},
 primaryClass = {astro-ph.SR},
       adsurl = {https://ui.adsabs.harvard.edu/abs/2016Natur.533..217R}
}

@ARTICLE{2017ApJ...844...90B,
       author = {{Berghea}, C.~T. and {Nelson}, George J. and {Rusu}, C.~E. and {Keeton}, C.~R. and {Dudik}, R.~P.},
        title = "{Discovery of the First Quadruple Gravitationally Lensed Quasar Candidate with Pan-STARRS}",
      journal = {\apj},
         year = 2017,
        month = aug,
       volume = {844},
       number = {2},
          eid = {90},
        pages = {90},
          doi = {10.3847/1538-4357/aa7aa6},
archivePrefix = {arXiv},
       eprint = {1705.08359},
 primaryClass = {astro-ph.GA},
       adsurl = {https://ui.adsabs.harvard.edu/abs/2017ApJ...844...90B}
}

@ARTICLE{2017A&A...605L...8L,
       author = {{Lee}, C.-H.},
        title = "{Accurate spectroscopic redshift of the multiply lensed quasar PSOJ0147 from the Pan-STARRS survey}",
      journal = {\aap},
         year = 2017,
        month = sep,
       volume = {605},
          eid = {L8},
        pages = {L8},
          doi = {10.1051/0004-6361/201731695},
archivePrefix = {arXiv},
       eprint = {1708.05131},
 primaryClass = {astro-ph.GA},
       adsurl = {https://ui.adsabs.harvard.edu/abs/2017A&A...605L...8L}
}

@ARTICLE{2018MNRAS.475.3086L,
       author = {{Lee}, Chien-Hsiu},
        title = "{A closer look at the quadruply lensed quasar PSOJ0147: spectroscopic redshifts and microlensing effect}",
      journal = {\mnras},
         year = 2018,
        month = apr,
       volume = {475},
       number = {3},
        pages = {3086-3089},
          doi = {10.1093/mnras/sty078},
archivePrefix = {arXiv},
       eprint = {1801.01851},
 primaryClass = {astro-ph.GA},
       adsurl = {https://ui.adsabs.harvard.edu/abs/2018MNRAS.475.3086L}
}

@ARTICLE{2017ApJ...836L..18M,
       author = {{Mediavilla}, E. and {Jim{\'e}nez-Vicente}, J. and {Mu{\~n}oz}, J.~A. and {Vives-Arias}, H. and {Calder{\'o}n-Infante}, J.},
        title = "{Limits on the Mass and Abundance of Primordial Black Holes from Quasar Gravitational Microlensing}",
      journal = {\apjl},
         year = 2017,
        month = feb,
       volume = {836},
       number = {2},
          eid = {L18},
        pages = {L18},
          doi = {10.3847/2041-8213/aa5dab},
archivePrefix = {arXiv},
       eprint = {1702.00947},
 primaryClass = {astro-ph.GA},
       adsurl = {https://ui.adsabs.harvard.edu/abs/2017ApJ...836L..18M}
}

@ARTICLE{2018MNRAS.480..245G,
       author = {{Guerin}, W. and {Rivet}, J.-P. and {Fouch{\'e}}, M. and {Labeyrie}, G. and {Vernet}, D. and {Vakili}, F. and {Kaiser}, R.},
        title = "{Spatial intensity interferometry on three bright stars}",
      journal = {\mnras},
         year = 2018,
        month = oct,
       volume = {480},
       number = {1},
        pages = {245-250},
          doi = {10.1093/mnras/sty1792},
archivePrefix = {arXiv},
       eprint = {1805.06653},
 primaryClass = {astro-ph.IM},
       adsurl = {https://ui.adsabs.harvard.edu/abs/2018MNRAS.480..245G}
}

@ARTICLE{2018NatAs...2..334K,
       author = {{Kelly}, Patrick L. and {Diego}, Jose M. and {Rodney}, Steven and {Kaiser}, Nick and {Broadhurst}, Tom and {Zitrin}, Adi and {Treu}, Tommaso and {P{\'e}rez-Gonz{\'a}lez}, Pablo G. and {Morishita}, Takahiro and {Jauzac}, Mathilde and {Selsing}, Jonatan and {Oguri}, Masamune and {Pueyo}, Laurent and {Ross}, Timothy W. and {Filippenko}, Alexei V. and {Smith}, Nathan and {Hjorth}, Jens and {Cenko}, S. Bradley and {Wang}, Xin and {Howell}, D. Andrew and {Richard}, Johan and {Frye}, Brenda L. and {Jha}, Saurabh W. and {Foley}, Ryan J. and {Norman}, Colin and {Bradac}, Marusa and {Zheng}, Weikang and {Brammer}, Gabriel and {Benito}, Alberto Molino and {Cava}, Antonio and {Christensen}, Lise and {de Mink}, Selma E. and {Graur}, Or and {Grillo}, Claudio and {Kawamata}, Ryota and {Kneib}, Jean-Paul and {Matheson}, Thomas and {McCully}, Curtis and {Nonino}, Mario and {P{\'e}rez-Fournon}, Ismael and {Riess}, Adam G. and {Rosati}, Piero and {Schmidt}, Kasper Borello and {Sharon}, Keren and {Weiner}, Benjamin J.},
        title = "{Extreme magnification of an individual star at redshift 1.5 by a galaxy-cluster lens}",
      journal = {Nature Astronomy},
         year = 2018,
        month = apr,
       volume = {2},
        pages = {334-342},
          doi = {10.1038/s41550-018-0430-3},
archivePrefix = {arXiv},
       eprint = {1706.10279},
 primaryClass = {astro-ph.GA},
       adsurl = {https://ui.adsabs.harvard.edu/abs/2018NatAs...2..334K}
}

@ARTICLE{2018PhRvD..97b3518O,
       author = {{Oguri}, Masamune and {Diego}, Jose M. and {Kaiser}, Nick and {Kelly}, Patrick L. and {Broadhurst}, Tom},
        title = "{Understanding caustic crossings in giant arcs: Characteristic scales, event rates, and constraints on compact dark matter}",
      journal = {\prd},
         year = 2018,
        month = jan,
       volume = {97},
       number = {2},
          eid = {023518},
        pages = {023518},
          doi = {10.1103/PhysRevD.97.023518},
archivePrefix = {arXiv},
       eprint = {1710.00148},
 primaryClass = {astro-ph.CO},
       adsurl = {https://ui.adsabs.harvard.edu/abs/2018PhRvD..97b3518O}
}

@ARTICLE{2018MNRAS.478.4816S,
       author = {{Spingola}, C. and {McKean}, J.~P. and {Auger}, M.~W. and {Fassnacht}, C.~D. and {Koopmans}, L.~V.~E. and {Lagattuta}, D.~J. and {Vegetti}, S.},
        title = "{SHARP - V. Modelling gravitationally lensed radio arcs imaged with global VLBI observations}",
      journal = {\mnras},
         year = 2018,
        month = aug,
       volume = {478},
       number = {4},
        pages = {4816-4829},
          doi = {10.1093/mnras/sty1326},
archivePrefix = {arXiv},
       eprint = {1807.05566},
 primaryClass = {astro-ph.GA},
       adsurl = {https://ui.adsabs.harvard.edu/abs/2018MNRAS.478.4816S}
}

@ARTICLE{2019MNRAS.485.3009H,
       author = {{Hartley}, P. and {Jackson}, N. and {Sluse}, D. and {Stacey}, H.~R. and {Vives-Arias}, H.},
        title = "{Strong lensing reveals jets in a sub-microJy radio-quiet quasar}",
      journal = {\mnras},
         year = 2019,
        month = may,
       volume = {485},
       number = {3},
        pages = {3009-3023},
          doi = {10.1093/mnras/stz510},
archivePrefix = {arXiv},
       eprint = {1901.05791},
 primaryClass = {astro-ph.GA},
       adsurl = {https://ui.adsabs.harvard.edu/abs/2019MNRAS.485.3009H}
}

@ARTICLE{2019NatAs...3..524N,
       author = {{Niikura}, Hiroko and {Takada}, Masahiro and {Yasuda}, Naoki and {Lupton}, Robert H. and {Sumi}, Takahiro and {More}, Surhud and {Kurita}, Toshiki and {Sugiyama}, Sunao and {More}, Anupreeta and {Oguri}, Masamune and {Chiba}, Masashi},
        title = "{Microlensing constraints on primordial black holes with Subaru/HSC Andromeda observations}",
      journal = {Nature Astronomy},
         year = 2019,
        month = apr,
       volume = {3},
        pages = {524-534},
          doi = {10.1038/s41550-019-0723-1},
archivePrefix = {arXiv},
       eprint = {1701.02151},
 primaryClass = {astro-ph.CO},
       adsurl = {https://ui.adsabs.harvard.edu/abs/2019NatAs...3..524N}
}

@ARTICLE{2019MNRAS.483.5649S,
       author = {{Shajib}, A.~J. and {Birrer}, S. and {Treu}, T. and {Auger}, M.~W. and {Agnello}, A. and {Anguita}, T. and {Buckley-Geer}, E.~J. and {Chan}, J.~H.~H. and {Collett}, T.~E. and {Courbin}, F. and {Fassnacht}, C.~D. and {Frieman}, J. and {Kayo}, I. and {Lemon}, C. and {Lin}, H. and {Marshall}, P.~J. and {McMahon}, R. and {More}, A. and {Morgan}, N.~D. and {Motta}, V. and {Oguri}, M. and {Ostrovski}, F. and {Rusu}, C.~E. and {Schechter}, P.~L. and {Shanks}, T. and {Suyu}, S.~H. and {Meylan}, G. and {Abbott}, T.~M.~C. and {Allam}, S. and {Annis}, J. and {Avila}, S. and {Bertin}, E. and {Brooks}, D. and {Carnero Rosell}, A. and {Carrasco Kind}, M. and {Carretero}, J. and {Cunha}, C.~E. and {da Costa}, L.~N. and {De Vicente}, J. and {Desai}, S. and {Doel}, P. and {Flaugher}, B. and {Fosalba}, P. and {Garc{\'\i}a-Bellido}, J. and {Gerdes}, D.~W. and {Gruen}, D. and {Gruendl}, R.~A. and {Gutierrez}, G. and {Hartley}, W.~G. and {Hollowood}, D.~L. and {Hoyle}, B. and {James}, D.~J. and {Kuehn}, K. and {Kuropatkin}, N. and {Lahav}, O. and {Lima}, M. and {Maia}, M.~A.~G. and {March}, M. and {Marshall}, J.~L. and {Melchior}, P. and {Menanteau}, F. and {Miquel}, R. and {Plazas}, A.~A. and {Sanchez}, E. and {Scarpine}, V. and {Sevilla-Noarbe}, I. and {Smith}, M. and {Soares-Santos}, M. and {Sobreira}, F. and {Suchyta}, E. and {Swanson}, M.~E.~C. and {Tarle}, G. and {Walker}, A.~R.},
        title = "{Is every strong lens model unhappy in its own way? Uniform modelling of a sample of 13 quadruply+ imaged quasars}",
      journal = {\mnras},
         year = 2019,
        month = mar,
       volume = {483},
       number = {4},
        pages = {5649-5671},
          doi = {10.1093/mnras/sty3397},
archivePrefix = {arXiv},
       eprint = {1807.09278},
 primaryClass = {astro-ph.GA},
       adsurl = {https://ui.adsabs.harvard.edu/abs/2019MNRAS.483.5649S}
}

@ARTICLE{2019MNRAS.483.5583V,
       author = {{Vernardos}, G.},
        title = "{Microlensing flux ratio predictions for Euclid}",
      journal = {\mnras},
         year = 2019,
        month = mar,
       volume = {483},
       number = {4},
        pages = {5583-5594},
          doi = {10.1093/mnras/sty3486},
archivePrefix = {arXiv},
       eprint = {1901.01246},
 primaryClass = {astro-ph.GA},
       adsurl = {https://ui.adsabs.harvard.edu/abs/2019MNRAS.483.5583V}
}

@ARTICLE{2020MNRAS.492..742A,
       author = {{Asadi}, Saghar and {Zackrisson}, Erik and {Varenius}, Eskil and {Freeland}, Emily and {Conway}, John and {Wiik}, Kaj},
        title = "{The case against gravitational millilensing in the multiply-imaged quasar B1152+199}",
      journal = {\mnras},
         year = 2020,
        month = feb,
       volume = {492},
       number = {1},
        pages = {742-748},
          doi = {10.1093/mnras/stz3450},
archivePrefix = {arXiv},
       eprint = {1811.06053},
 primaryClass = {astro-ph.GA},
       adsurl = {https://ui.adsabs.harvard.edu/abs/2020MNRAS.492..742A}
}

@ARTICLE{2020NaPho..14..250K,
       author = {{Korzh}, Boris and {Zhao}, Qing-Yuan and {Allmaras}, Jason P. and {Frasca}, Simone and {Autry}, Travis M. and {Bersin}, Eric A. and {Beyer}, Andrew D. and {Briggs}, Ryan M. and {Bumble}, Bruce and {Colangelo}, Marco and {Crouch}, Garrison M. and {Dane}, Andrew E. and {Gerrits}, Thomas and {Lita}, Adriana E. and {Marsili}, Francesco and {Moody}, Galan and {Pe{\~n}a}, Cristi{\'a}n and {Ramirez}, Edward and {Rezac}, Jake D. and {Sinclair}, Neil and {Stevens}, Martin J. and {Velasco}, Angel E. and {Verma}, Varun B. and {Wollman}, Emma E. and {Xie}, Si and {Zhu}, Di and {Hale}, Paul D. and {Spiropulu}, Maria and {Silverman}, Kevin L. and {Mirin}, Richard P. and {Nam}, Sae Woo and {Kozorezov}, Alexander G. and {Shaw}, Matthew D. and {Berggren}, Karl K.},
        title = "{Demonstration of sub-3 ps temporal resolution with a superconducting nanowire single-photon detector}",
      journal = {Nature Photonics},
         year = 2020,
        month = apr,
       volume = {14},
       number = {4},
        pages = {250-255},
          doi = {10.1038/s41566-020-0589-x},
archivePrefix = {arXiv},
       eprint = {1804.06839},
 primaryClass = {physics.ins-det},
       adsurl = {https://ui.adsabs.harvard.edu/abs/2020NaPho..14..250K}
}

@article{Reddy:20,
author = {Dileep V. Reddy and Robert R. Nerem and Sae Woo Nam and Richard P. Mirin and Varun B. Verma},
journal = {Optica},
number = {12},
pages = {1649--1653},
publisher = {Optica Publishing Group},
title = {Superconducting nanowire single-photon detectors with 98\% system detection efficiency at 1550  nm},
volume = {7},
month = {Dec},
year = {2020},
url = {https://opg.optica.org/optica/abstract.cfm?URI=optica-7-12-1649},
doi = {10.1364/OPTICA.400751},
}

@ARTICLE{2020MNRAS.494..218R,
       author = {{Rivet}, J.-P. and {Siciak}, A. and {de Almeida}, E.~S.~G. and {Vakili}, F. and {Domiciano de Souza}, A. and {Fouch{\'e}}, M. and {Lai}, O. and {Vernet}, D. and {Kaiser}, R. and {Guerin}, W.},
        title = "{Intensity interferometry of P Cygni in the H {\ensuremath{\alpha}} emission line: towards distance calibration of LBV supergiant stars}",
      journal = {\mnras},
         year = 2020,
        month = may,
       volume = {494},
       number = {1},
        pages = {218-227},
          doi = {10.1093/mnras/staa588},
archivePrefix = {arXiv},
       eprint = {1910.08366},
 primaryClass = {astro-ph.IM},
       adsurl = {https://ui.adsabs.harvard.edu/abs/2020MNRAS.494..218R}
}

@ARTICLE{2020MNRAS.498.1420W,
       author = {{Wong}, Kenneth C. and {Suyu}, Sherry H. and {Chen}, Geoff C.-F. and {Rusu}, Cristian E. and {Millon}, Martin and {Sluse}, Dominique and {Bonvin}, Vivien and {Fassnacht}, Christopher D. and {Taubenberger}, Stefan and {Auger}, Matthew W. and {Birrer}, Simon and {Chan}, James H.~H. and {Courbin}, Frederic and {Hilbert}, Stefan and {Tihhonova}, Olga and {Treu}, Tommaso and {Agnello}, Adriano and {Ding}, Xuheng and {Jee}, Inh and {Komatsu}, Eiichiro and {Shajib}, Anowar J. and {Sonnenfeld}, Alessandro and {Blandford}, Roger D. and {Koopmans}, L{\'e}on V.~E. and {Marshall}, Philip J. and {Meylan}, Georges},
        title = "{H0LiCOW - XIII. A 2.4 per cent measurement of H$_{0}$ from lensed quasars: 5.3{\ensuremath{\sigma}} tension between early- and late-Universe probes}",
      journal = {\mnras},
         year = 2020,
        month = oct,
       volume = {498},
       number = {1},
        pages = {1420-1439},
          doi = {10.1093/mnras/stz3094},
archivePrefix = {arXiv},
       eprint = {1907.04869},
 primaryClass = {astro-ph.CO},
       adsurl = {https://ui.adsabs.harvard.edu/abs/2020MNRAS.498.1420W}
}

@article{Gramuglia:2022oii,
    author = "Gramuglia, Francesco and Wu, Ming-Lo and Bruschini, Claudio and Lee, Myung-Jae and Charbon, Edoardo",
    title = "{A Low-Noise CMOS SPAD Pixel With 12.1 Ps SPTR and 3 Ns Dead Time}",
    doi = "10.1109/JSTQE.2021.3088216",
    journal = "IEEE J. Sel. Top. Quant. Electron.",
    volume = "28",
    number = "2",
    pages = "3800809",
    year = "2022"
}

@ARTICLE{2022Natur.603..815W,
       author = {{Welch}, Brian and {Coe}, Dan and {Diego}, Jose M. and {Zitrin}, Adi and {Zackrisson}, Erik and {Dimauro}, Paola and {Jim{\'e}nez-Teja}, Yolanda and {Kelly}, Patrick and {Mahler}, Guillaume and {Oguri}, Masamune and {Timmes}, F.~X. and {Windhorst}, Rogier and {Florian}, Michael and {de Mink}, S.~E. and {Avila}, Roberto J. and {Anderson}, Jay and {Bradley}, Larry and {Sharon}, Keren and {Vikaeus}, Anton and {McCandliss}, Stephan and {Brada{\v{c}}}, Maru{\v{s}}a and {Rigby}, Jane and {Frye}, Brenda and {Toft}, Sune and {Strait}, Victoria and {Trenti}, Michele and {Sharma}, Soniya and {Andrade-Santos}, Felipe and {Broadhurst}, Tom},
        title = "{A highly magnified star at redshift 6.2}",
      journal = {\nat},
         year = 2022,
        month = mar,
       volume = {603},
       number = {7903},
        pages = {815-818},
          doi = {10.1038/s41586-022-04449-y},
archivePrefix = {arXiv},
       eprint = {2209.14866},
 primaryClass = {astro-ph.GA},
       adsurl = {https://ui.adsabs.harvard.edu/abs/2022Natur.603..815W}
}

@ARTICLE{2023ApJ...944L...6M,
       author = {{Meena}, Ashish Kumar and {Zitrin}, Adi and {Jim{\'e}nez-Teja}, Yolanda and {Zackrisson}, Erik and {Chen}, Wenlei and {Coe}, Dan and {Diego}, Jose M. and {Dimauro}, Paola and {Furtak}, Lukas J. and {Kelly}, Patrick L. and {Oguri}, Masamune and {Welch}, Brian and {Abdurro'uf} and {Andrade-Santos}, Felipe and {Adamo}, Angela and {Bhatawdekar}, Rachana and {Brada{\v{c}}}, Maru{\v{s}}a and {Bradley}, Larry D. and {Broadhurst}, Tom and {Conselice}, Christopher J. and {Dayal}, Pratika and {Donahue}, Megan and {Frye}, Brenda L. and {Fujimoto}, Seiji and {Hsiao}, Tiger Yu-Yang and {Kokorev}, Vasily and {Mahler}, Guillaume and {Vanzella}, Eros and {Windhorst}, Rogier A.},
        title = "{Two Lensed Star Candidates at z ≃ 4.8 behind the Galaxy Cluster MACS J0647.7+7015}",
      journal = {\apjl},
         year = 2023,
        month = feb,
       volume = {944},
       number = {1},
          eid = {L6},
        pages = {L6},
          doi = {10.3847/2041-8213/acb645},
archivePrefix = {arXiv},
       eprint = {2211.13334},
 primaryClass = {astro-ph.GA},
       adsurl = {https://ui.adsabs.harvard.edu/abs/2023ApJ...944L...6M}
}

@ARTICLE{2020MNRAS.498.4577B,
       author = {{Baumgartner}, Sandra and {Bernardini}, Mauro and {Canivete Cuissa}, Jos{\'e} R. and {de Laroussilhe}, Hugues and {Mitchell}, Alison M.~W. and {Neuenschwander}, Benno A. and {Saha}, Prasenjit and {Schaeffer}, Timoth{\'e}e and {Soyuer}, Deniz and {Zwick}, Lorenz},
        title = "{Towards a polarization prediction for LISA via intensity interferometry}",
      journal = {\mnras},
         year = 2020,
        month = nov,
       volume = {498},
       number = {3},
        pages = {4577-4589},
          doi = {10.1093/mnras/staa2638},
archivePrefix = {arXiv},
       eprint = {2008.11538},
 primaryClass = {astro-ph.IM},
       adsurl = {https://ui.adsabs.harvard.edu/abs/2020MNRAS.498.4577B}
}

@ARTICLE{2023AJ....165..117M,
       author = {{Matthews}, Nolan and {Rivet}, Jean-Pierre and {Vernet}, David and {Hugbart}, Mathilde and {Labeyrie}, Guillaume and {Kaiser}, Robin and {Chab{\'e}}, Julien and {Courde}, Cl{\'e}ment and {Lai}, Olivier and {Vakili}, Farrokh and {Garde}, Olivier and {Guerin}, William},
        title = "{Intensity Interferometry Observations of the H{\ensuremath{\alpha}} Envelope of {\ensuremath{\gamma}}Cas with M{\'e}O and a Portable Telescope}",
      journal = {\aj},
         year = 2023,
        month = mar,
       volume = {165},
       number = {3},
          eid = {117},
        pages = {117},
          doi = {10.3847/1538-3881/acb142},
archivePrefix = {arXiv},
       eprint = {2301.04878},
 primaryClass = {astro-ph.IM},
       adsurl = {https://ui.adsabs.harvard.edu/abs/2023AJ....165..117M}
}

@ARTICLE{2023ConPh..64...47P,
       author = {{Padovani}, Paolo and {Cirasuolo}, Michele},
        title = "{The Extremely Large Telescope}",
      journal = {Contemporary Physics},
         year = 2023,
        month = jan,
       volume = {64},
       number = {1},
        pages = {47-64},
          doi = {10.1080/00107514.2023.2266921},
archivePrefix = {arXiv},
       eprint = {2312.04299},
 primaryClass = {astro-ph.IM},
       adsurl = {https://ui.adsabs.harvard.edu/abs/2023ConPh..64...47P}
}

@article{Walter:2023yF,
  author = {{Walter}, R. and  {Charbon}, E. and {della Volpe}, D. and  {Lyard}, E. and {Produit}, N. and {Saha}, P. and {Sliusar}, V.  and  Tramacere, A.},
  title = "{Resolving accretion disks with quantum optics}",
  doi = "10.22323/1.444.1491",
  journal = "PoS",
  year = 2023,
  volume = "ICRC2023",
  pages = "1491"
}

@ARTICLE{2024MNRAS.529.4387A,
       author = {{Abe}, S. and {Abhir}, J. and {Acciari}, V.~A. and {Aguasca-Cabot}, A. and {Agudo}, I. and {Aniello}, T. and {Ansoldi}, S. and {Antonelli}, L.~A. and {Arbet Engels}, A. and {Arcaro}, C. and {Artero}, M. and {Asano}, K. and {Babi{\'c}}, A. and {Baquero}, A. and {de Almeida}, U. Barres and {Barrio}, J.~A. and {Batkovi{\'c}}, I. and {Bautista}, A. and {Baxter}, J. and {Gonz{\'a}lez}, J. Becerra and {Bernardini}, E. and {Bernardos}, M. and {Bernete}, J. and {Berti}, A. and {Besenrieder}, J. and {Bigongiari}, C. and {Biland}, A. and {Blanch}, O. and {Bonnoli}, G. and {Bo{\v{s}}njak}, {\v{Z}}. and {Burelli}, I. and {Busetto}, G. and {Campoy-Ordaz}, A. and {Carosi}, A. and {Carosi}, R. and {Carretero-Castrillo}, M. and {Ceribella}, G. and {Chai}, Y. and {Cifuentes}, A. and {Colombo}, E. and {Contreras}, J.~L. and {Cortina}, J. and {Covino}, S. and {D'Amico}, G. and {D'Elia}, V. and {Da Vela}, P. and {Dazzi}, F. and {De Angelis}, A. and {De Lotto}, B. and {de Menezes}, R. and {Del Popolo}, A. and {Delfino}, M. and {Delgado}, J. and {Delgado Mendez}, C. and {Di Pierro}, F. and {Di Venere}, L. and {Dominis Prester}, D. and {Donini}, A. and {Dorner}, D. and {Doro}, M. and {Elsaesser}, D. and {Emery}, G. and {Escudero}, J. and {Fari{\~n}a}, L. and {Fattorini}, A. and {Foffano}, L. and {Font}, L. and {Fr{\"o}se}, S. and {Fukami}, S. and {Fukazawa}, Y. and {Garc{\'\i}a L{\'o}pez}, R.~J. and {Garczarczyk}, M. and {Gasparyan}, S. and {Gaug}, M. and {Giesbrecht Paiva}, J.~G. and {Giglietto}, N. and {Giordano}, F. and {Gliwny}, P. and {Gradetzke}, T. and {Grau}, R. and {Green}, D. and {Green}, J.~G. and {G{\"u}nther}, P. and {Hadasch}, D. and {Hahn}, A. and {Hassan}, T. and {Heckmann}, L. and {Herrera}, J. and {Hrupec}, D. and {H{\"u}tten}, M. and {Imazawa}, R. and {Ishio}, K. and {Jim{\'e}nez Mart{\'\i}nez}, I. and {Jormanainen}, J. and {Kayanoki}, T. and {Kerszberg}, D. and {Kluge}, G.~W. and {Kobayashi}, Y. and {Kouch}, P.~M. and {Kubo}, H. and {Kushida}, J. and {L{\'a}inez}, M. and {Lamastra}, A. and {Leone}, F. and {Lindfors}, E. and {Linhoff}, L. and {Lombardi}, S. and {Longo}, F. and {L{\'o}pez-Coto}, R. and {L{\'o}pez-Moya}, M. and {L{\'o}pez-Oramas}, A. and {Loporchio}, S. and {Lorini}, A. and {Lyard}, E. and {Machado de Oliveira Fraga}, B. and {Majumdar}, P. and {Makariev}, M. and {Maneva}, G. and {Mang}, N. and {Manganaro}, M. and {Mangano}, S. and {Mannheim}, K. and {Mariotti}, M. and {Mart{\'\i}nez}, M. and {Mart{\'\i}nez-Chicharro}, M. and {Mas-Aguilar}, A. and {Mazin}, D. and {Menchiari}, S. and {Mender}, S. and {Miceli}, D. and {Miener}, T. and {Miranda}, J.~M. and {Mirzoyan}, R. and {Molero Gonz{\'a}lez}, M. and {Molina}, E. and {Mondal}, H.~A. and {Moralejo}, A. and {Morcuende}, D. and {Nakamori}, T. and {Nanci}, C. and {Neustroev}, V. and {Nickel}, L. and {Nievas Rosillo}, M. and {Nigro}, C. and {Nikoli{\'c}}, L. and {Nilsson}, K. and {Nishijima}, K. and {Ekoume}, T. Njoh and {Noda}, K. and {Nozaki}, S. and {Ohtani}, Y. and {Okumura}, A. and {Otero-Santos}, J. and {Paiano}, S. and {Palatiello}, M. and {Paneque}, D. and {Paoletti}, R. and {Paredes}, J.~M. and {Peresano}, M. and {Persic}, M. and {Pihet}, M. and {Pirola}, G. and {Podobnik}, F. and {Prada Moroni}, P.~G. and {Prandini}, E. and {Principe}, G. and {Priyadarshi}, C. and {Rhode}, W. and {Rib{\'o}}, M. and {Rico}, J. and {Righi}, C. and {Sahakyan}, N. and {Saito}, T. and {Satalecka}, K. and {Saturni}, F.~G. and {Schleicher}, B. and {Schmidt}, K. and {Schmuckermaier}, F. and {Schubert}, J.~L. and {Schweizer}, T. and {Sciaccaluga}, A. and {Silvestri}, G. and {Sitarek}, J. and {Sliusar}, V. and {Sobczynska}, D. and {Spolon}, A. and {Stamerra}, A. and {Stri{\v{s}}kovi{\'c}}, J. and {Strom}, D. and {Strzys}, M. and {Suda}, Y. and {Suri{\'c}}, T. and {Suutarinen}, S. and {Tajima}, H. and {Takahashi}, M. and {Takeishi}, R. and {Temnikov}, P. and {Terauchi}, K. and {Terzi{\'c}}, T. and {Teshima}, M.},
        title = "{Performance and first measurements of the MAGIC stellar intensity interferometer}",
      journal = {\mnras},
         year = 2024,
        month = apr,
       volume = {529},
       number = {4},
        pages = {4387-4404},
          doi = {10.1093/mnras/stae697},
archivePrefix = {arXiv},
       eprint = {2402.04755},
 primaryClass = {astro-ph.IM},
       adsurl = {https://ui.adsabs.harvard.edu/abs/2024MNRAS.529.4387A}
}

@ARTICLE{2024NatCo..15.3973C,
       author = {{Charaev}, Ilya and {Batson}, Emma K. and {Cherednichenko}, Sergey and {Reidy}, Kate and {Drakinskiy}, Vladimir and {Yu}, Yang and {Lara-Avila}, Samuel and {Thomsen}, Joachim D. and {Colangelo}, Marco and {Incalza}, Francesca and {Ilin}, Konstantin and {Schilling}, Andreas and {Berggren}, Karl K.},
        title = "{Single-photon detection using large-scale high-temperature MgB$_{2}$ sensors at 20 K}",
      journal = {Nature Communications},
         year = 2024,
        month = may,
       volume = {15},
          eid = {3973},
        pages = {3973},
          doi = {10.1038/s41467-024-47353-x},
archivePrefix = {arXiv},
       eprint = {2308.15228},
 primaryClass = {cond-mat.supr-con},
       adsurl = {https://ui.adsabs.harvard.edu/abs/2024NatCo..15.3973C}
}

@ARTICLE{2024PhRvD.109l3029D,
       author = {{Dalal}, Neal and {Galanis}, Marios and {Gammie}, Charles and {Gralla}, Samuel E. and {Murray}, Norman},
        title = "{Probing H$_{0}$ and resolving AGN disks with ultrafast photon counters}",
      journal = {\prd},
         year = 2024,
        month = jun,
       volume = {109},
       number = {12},
          eid = {123029},
        pages = {123029},
          doi = {10.1103/PhysRevD.109.123029},
archivePrefix = {arXiv},
       eprint = {2403.15903},
 primaryClass = {astro-ph.CO},
       adsurl = {https://ui.adsabs.harvard.edu/abs/2024PhRvD.109l3029D}
}

@ARTICLE{2025ApJ...995..191A,
       author = {{Archer}, A. and {Aufdenberg}, J.~P. and {Bangale}, P. and {Bartkoske}, J.~T. and {Benbow}, W. and {Buckley}, J.~H. and {Chen}, Y. and {Chin}, N.~B.~Y. and {Christiansen}, J.~L. and {Chromey}, A.~J. and {Duerr}, A. and {Escobar Godoy}, M. and {Feldman}, S. and {Feng}, Q. and {Filbert}, S. and {Fortson}, L. and {Furniss}, A. and {Hanlon}, W. and {Hervet}, O. and {Hinrichs}, C.~E. and {Holder}, J. and {Hughes}, Z. and {Humensky}, T.~B. and {Jin}, W. and {Johnson}, M.~N. and {Kertzman}, M. and {Kherlakian}, M. and {Kieda}, D. and {Korzoun}, N. and {Lebohec}, T. and {Lisa}, M.~A. and {Lundy}, M. and {Maier}, G. and {Matthews}, N. and {Moriarty}, P. and {Mukherjee}, R. and {Ning}, W. and {Ong}, R.~A. and {Pandey}, A. and {Pohl}, M. and {Pueschel}, E. and {Quinn}, J. and {Rabinowitz}, P.~L. and {Ragan}, K. and {Reynolds}, P.~T. and {Ribeiro}, D. and {Roache}, E. and {Rose}, J.~G. and {Sadeh}, I. and {Saha}, L. and {Santander}, M. and {Scott}, J. and {Sembroski}, G.~H. and {Shang}, R. and {Tak}, D. and {Tucci}, J.~V. and {Valverde}, J. and {Vassiliev}, V.~V. and {Williams}, D.~A. and {Wong}, S.~L. and {VERITAS Collaboration}},
        title = "{Measurement of the Photosphere Oblateness of {\ensuremath{\gamma}} Cassiopeiae via Stellar Intensity Interferometry with the VERITAS Observatory}",
      journal = {\apj},
         year = 2025,
        month = dec,
       volume = {995},
       number = {2},
          eid = {191},
        pages = {191},
          doi = {10.3847/1538-4357/ae0744},
archivePrefix = {arXiv},
       eprint = {2506.15027},
 primaryClass = {astro-ph.SR},
       adsurl = {https://ui.adsabs.harvard.edu/abs/2025ApJ...995..191A}
}

@ARTICLE{2025ApJS..280...49M,
       author = {{Mr{\'o}z}, Przemek and {Udalski}, Andrzej and {Szyma{\'n}ski}, Micha{\l} K. and {Soszy{\'n}ski}, Igor and {Pietrukowicz}, Pawe{\l} and {Koz{\l}owski}, Szymon and {Poleski}, Rados{\l}aw and {Skowron}, Jan and {Skowron}, Dorota and {Ulaczyk}, Krzysztof and {Gromadzki}, Mariusz and {Rybicki}, Krzysztof and {Iwanek}, Patryk and {Wrona}, Marcin and {Ratajczak}, Milena},
        title = "{Microlensing Optical Depth, Event Rate, and Limits on Compact Objects in Dark Matter Based on 20 Yr of OGLE Observations of the Small Magellanic Cloud}",
      journal = {\apjs},
         year = 2025,
        month = oct,
       volume = {280},
       number = {2},
          eid = {49},
        pages = {49},
          doi = {10.3847/1538-4365/adf842},
archivePrefix = {arXiv},
       eprint = {2507.13794},
 primaryClass = {astro-ph.GA},
       adsurl = {https://ui.adsabs.harvard.edu/abs/2025ApJS..280...49M}
}

@ARTICLE{2025A&A...701A..35N,
       author = {{Neira}, F. and {Anguita}, T. and {Vernardos}, G.},
        title = "{Simulating quasar microlensing light curves: High magnification events}",
      journal = {\aap},
         year = 2025,
        month = sep,
       volume = {701},
          eid = {A35},
        pages = {A35},
          doi = {10.1051/0004-6361/202555125},
archivePrefix = {arXiv},
       eprint = {2507.21973},
 primaryClass = {astro-ph.GA},
       adsurl = {https://ui.adsabs.harvard.edu/abs/2025A&A...701A..35N}
}

@ARTICLE{2025ApJ...979...13P,
       author = {{Pascale}, Massimo and {Frye}, Brenda L. and {Pierel}, Justin D.~R. and {Chen}, Wenlei and {Kelly}, Patrick L. and {Cohen}, Seth H. and {Windhorst}, Rogier A. and {Riess}, Adam G. and {Kamieneski}, Patrick S. and {Diego}, Jos{\'e} M. and {Meena}, Ashish K. and {Cha}, Sangjun and {Oguri}, Masamune and {Zitrin}, Adi and {Jee}, M. James and {Foo}, Nicholas and {Leimbach}, Reagen and {Koekemoer}, Anton M. and {Conselice}, C.~J. and {Dai}, Liang and {Goobar}, Ariel and {Siebert}, Matthew R. and {Strolger}, Lou and {Willner}, S.~P.},
        title = "{SN H0pe: The First Measurement of H$_{0}$ from a Multiply Imaged Type Ia Supernova, Discovered by JWST}",
      journal = {\apj},
         year = 2025,
        month = jan,
       volume = {979},
       number = {1},
          eid = {13},
        pages = {13},
          doi = {10.3847/1538-4357/ad9928},
archivePrefix = {arXiv},
       eprint = {2403.18902},
 primaryClass = {astro-ph.CO},
       adsurl = {https://ui.adsabs.harvard.edu/abs/2025ApJ...979...13P}
}

@ARTICLE{2025Reson30.45R,
       author = {{Rai}, K.~N. and {Basak}, S. and {Sarangi}, S. and {Saha}, P.},
        title = "{Interference with (Pseudo) Thermal Light.}",
      journal = {Reson},
         year = 2025,
        month = feb,
       volume = {30},
        pages = {45-57},
          doi = {10.1007/s12045-025-1729-x}
}

@article{Saha:2025uG,
  author = "Saha, Prasenjit  and  Biteau, Jonathan  and  Carlile, Colin  and  Dravins, Dainis  and  Fiori, Michele  and  Hassan, Tarek  and  Kieda, Dave  and  Lisa, Mike  and  Luce, Quentin  and  Mitchell, Alison  and  Spolon, Alessia  and  Stanic, Lucijana  and  Vogel, Naomi  and  Zampieri, Luca  and  Zmija, Andreas",
  title = "{Intensity Interferometry prospects with the CTAO}",
  doi = "10.22323/1.501.0959",
  journal = "PoS",
  year = 2025,
  volume = "ICRC2025",
  pages = "959"
}

@ARTICLE{2025arXiv250912319S,
       author = {{Suyu}, S.~H. and {Acebron}, A. and {Grillo}, C. and {Bergamini}, P. and {Caminha}, G.~B. and {Cha}, S. and {Diego}, J.~M. and {Ertl}, S. and {Foo}, N. and {Frye}, B.~L. and {Fudamoto}, Y. and {Granata}, G. and {Halkola}, A. and {Jee}, M.~J. and {Kamieneski}, P.~S. and {Koekemoer}, A.~M. and {Meena}, A.~K. and {Newman}, A.~B. and {Nishida}, S. and {Oguri}, M. and {Rosati}, P. and {Schuldt}, S. and {Zitrin}, A. and {Ca{\~n}ameras}, R. and {Hayes}, E.~E. and {Larison}, C. and {Mamuzic}, E. and {Millon}, M. and {Pierel}, J.~D.~R. and {Tortorelli}, L. and {Wang}, H.},
        title = "{Cosmology with supernova Encore in the strong lensing cluster MACS J0138-2155: Lens model comparison and H0 measurement}",
      journal = {arXiv e-prints},
         year = 2025,
        month = sep,
          eid = {arXiv:2509.12319},
        pages = {arXiv:2509.12319},
          doi = {10.48550/arXiv.2509.12319},
archivePrefix = {arXiv},
       eprint = {2509.12319},
 primaryClass = {astro-ph.CO},
       adsurl = {https://ui.adsabs.harvard.edu/abs/2025arXiv250912319S}
}

@ARTICLE{2025MNRAS.537.2334V,
       author = {{Vogel}, Naomi and {Zmija}, Andreas and {Wohlleben}, Frederik and {Anton}, Gisela and {Mitchell}, Alison and {Zink}, Adrian and {Funk}, Stefan},
        title = "{Simultaneous two-colour intensity interferometry with H.E.S.S}",
      journal = {\mnras},
         year = 2025,
        month = mar,
       volume = {537},
       number = {3},
        pages = {2334-2341},
          doi = {10.1093/mnras/stae2643},
archivePrefix = {arXiv},
       eprint = {2411.16471},
 primaryClass = {astro-ph.IM},
       adsurl = {https://ui.adsabs.harvard.edu/abs/2025MNRAS.537.2334V}
}

@ARTICLE{2025arXiv251212470S,
       author = {{Sliusar}, Vitalii and {Della Volpe}, Domenico and {Garcia}, Benjamin and {Koziol}, Gilles and {Lyard}, Etienne and {Produit}, Nicolas and {Raiola}, Aramis and {Saha}, Prasenjit and {Stanic}, Lucijana and {Walter}, Roland},
        title = "{Exploiting light coherence in astrophysics}",
      journal = {arXiv e-prints},
         year = 2025,
        month = dec,
          eid = {arXiv:2512.12470},
        pages = {arXiv:2512.12470},
          doi = {10.48550/arXiv.2512.12470},
archivePrefix = {arXiv},
       eprint = {2512.12470},
 primaryClass = {astro-ph.GA},
       adsurl = {https://ui.adsabs.harvard.edu/abs/2025arXiv251212470S}
}

@ARTICLE{2026arXiv260212717K,
       author = {{Kaiser}, Robin and {Guerin}, William and {Vakili}, Farrokh and {Berger}, Jean-Philippe and {Nomerotski}, Andrei and {Kulkov}, Sergei and {Svihra}, Peter and {Santos}, Eva and {Carlile}, Colin and {Dravins}, Dainis and {Funk}, Stefan and {Saha}, Prasenjit and {Walter}, Roland and {Borges Fernandes}, Marcelo and {Kim}, Alex G. and {Dunsky}, David and {Van Tilburg}, Ken and {Baryakhtar}, Masha and {Galanis}, Marios and {Wagoner}, Robert V. and {Dalal}, Neal and {Huang}, Junwu and {Gammie}, Charles and {Murray}, Norman W.},
        title = "{ESO White Paper on Intensity Interferometry: Cosmology, Fundamental Physics, Quantum Optics}",
      journal = {arXiv e-prints},
         year = 2026,
        month = feb,
          eid = {arXiv:2602.12717},
        pages = {arXiv:2602.12717},
          doi = {10.48550/arXiv.2602.12717},
archivePrefix = {arXiv},
       eprint = {2602.12717},
 primaryClass = {astro-ph.IM},
       adsurl = {https://ui.adsabs.harvard.edu/abs/2026arXiv260212717K}
}

@ARTICLE{2026arXiv260413152D,
       author = {{Davies}, Frederick B. and {Ba{\~n}ados}, Eduardo and {Bosman}, Sarah E.~I. and {Ganguly}, Arpita and {Belladitta}, Silvia and {Power}, Jennifer and {Rees}, Jon},
        title = "{Persephone's Torch: A 15th Magnitude Quadruply-Lensed Quasar From the Couch Discovered with SPHEREx and the LBT}",
      journal = {arXiv e-prints},
         year = 2026,
        month = apr,
          eid = {arXiv:2604.13152},
        pages = {arXiv:2604.13152},
archivePrefix = {arXiv},
       eprint = {2604.13152},
 primaryClass = {astro-ph.GA},
       adsurl = {https://ui.adsabs.harvard.edu/abs/2026arXiv260413152D}
}

\end{document}